\pdfoutput=1

\documentclass[journal,compsoc]{IEEEtran}
\usepackage{framed}
\usepackage{listings}
\usepackage{xcolor}
\usepackage{color}
\usepackage[]{collab}
\usepackage{multirow}
\usepackage{algorithm}
\usepackage{algorithmic}
\usepackage{amsmath}
\usepackage{amssymb}
\usepackage{graphicx}
\usepackage{booktabs}
\usepackage{tcolorbox}
\usepackage{url}
\usepackage{ulem}
\usepackage{subfigure}
\usepackage{colortbl}
\usepackage{framed}
\definecolor{Gray}{gray}{0.9}

\definecolor{codegreen}{rgb}{0,0.6,0}
\definecolor{codegray}{rgb}{0.5,0.5,0.5}
\definecolor{codepurple}{rgb}{0.58,0,0.82}
\definecolor{backcolour}{rgb}{0.95,0.95,0.92}

\definecolor{mygray}{gray}{.9}

\lstdefinestyle{mystyle}{
    commentstyle=\color{orange},
    keywordstyle=\color{magenta},
    numberstyle=\tiny\color{codegray},
    stringstyle=\color{codepurple},
    basicstyle=\ttfamily\footnotesize,
    breakatwhitespace=false,         
    breaklines=true,                 
    captionpos=b,                    
    keepspaces=true,                 
    numbers=left,                    
    numbersep=5pt,                  
    showspaces=false,                
    showstringspaces=false,
    showtabs=false,                  
    tabsize=2
}

\newcommand{\bench}{\textsc{APIBench}\xspace}
\newcommand{\benchq}{\textsc{APIBench-Q}\xspace}
\newcommand{\benchc}{\textsc{APIBench-C}\xspace}

\newcommand\etal{{\it{et al.\ }}}

\newcommand{\yun}[1]{\textcolor{blue}{{#1}}}

\newcommand{\tabincell}[2]{\begin{tabular}{@{}#1@{}}#2\end{tabular}}
\newcommand{\eat}[1]{\if 0 #1 \fi}

\newcommand{\finding}[2]{
\begin{tcolorbox}[width=\linewidth,boxrule=0pt,top=1pt, bottom=1pt, left=1pt,right=1pt, colback=gray!20,colframe=gray!20]
\textbf{Finding #1:} \textit{#2}
\end{tcolorbox}}

\newcommand{\implication}[2]{
\begin{framed}
\noindent\textbf{Implication #1:} \textit{#2}
\end{framed}
}

\usepackage{pifont}

\newcommand{\g}{\cellcolor{mygray}}

\newcommand{\example}[5]{\begin{table}[h]
    \centering
    \begin{tabular}{lp{5.7cm}}
    \toprule
    \rowcolor{mygray}
    \multicolumn{2}{l}{\textbf{Example #1: #2}} \\
    \textsc{Technique} & #3 \\
    \textsc{Original Query} & #4 \\
    \textsc{Processed Query} & #5  \\
    \bottomrule
    \end{tabular}
\end{table}
}

\ifCLASSOPTIONcompsoc
  \usepackage[nocompress]{cite}
\else
  \usepackage{cite}
\fi
\ifCLASSINFOpdf
\else
\fi
\begin{document}
%
\title{Revisiting, Benchmarking and Exploring API Recommendation: How Far Are We?}


\author{\IEEEauthorblockN{Yun Peng\IEEEauthorrefmark{2},
Shuqing Li\IEEEauthorrefmark{2},
Wenwei Gu\IEEEauthorrefmark{2},
Yichen Li\IEEEauthorrefmark{2},
Wenxuan Wang\IEEEauthorrefmark{2},
Cuiyun Gao\thanks{* corresponding author.}\IEEEauthorrefmark{1}\IEEEauthorrefmark{3},
and Michael Lyu\IEEEauthorrefmark{2}
} \\
\IEEEauthorblockA{\IEEEauthorrefmark{2}The Chinese University of Hong Kong, Hong Kong, China\\
\IEEEauthorrefmark{3}Harbin Institute of Technology, Shenzhen, China} \\
\IEEEauthorblockA{\{ypeng, sqli21, wwgu21, ycli21, wxwang, lyu\}@cse.cuhk.edu.hk, gaocuiyun@hit.edu.cn}
}


%




\IEEEtitleabstractindextext{%
\begin{abstract}
Application Programming Interfaces (APIs), which encapsulate the implementation of specific functions as interfaces, greatly improve the efficiency of modern software development. As numbers of APIs spring up nowadays, developers can hardly be familiar with all the APIs and
usually need to search for appropriate APIs for usage. So lots of efforts have been devoted to improving the API recommendation task. However, it has been increasingly difficult to gauge the performance of new models due to the lack of a uniform definition of the task and a standardized benchmark. For example, some studies regard the task as a code completion problem, while others recommend relative APIs given natural language queries. To reduce the challenges and better facilitate future research, in this paper, we revisit the API recommendation task and aim at benchmarking the approaches. Specifically, the paper groups the approaches into two categories according to the task definition, i.e., query-based API recommendation and code-based API recommendation. We study 11 recently-proposed approaches along with 4 widely-used IDEs. One benchmark named as APIBench is then built for the two respective categories of approaches. Based on APIBench, we distill some actionable insights and challenges for API recommendation. We also achieve some implications and directions for improving the performance of recommending APIs, including data source selection, appropriate query reformulation, low resource setting, and
cross-domain adaptation.
\end{abstract}

}

\maketitle


%
\IEEEpeerreviewmaketitle

\section{Introduction}\label{sec:intro}

Application Programming Interfaces (APIs) provided by software libraries or frameworks play an important role in modern software development. Almost all programs, even the basic ``hello world!" program, include at least one API. However, there are a huge number of APIs from different modules or libraries. For example, Java standard library~\cite{javadoc} provides more than 30,000 APIs. It is infeasible for developers to be familiar with all APIs. To address this problem, many approaches are proposed to recommend APIs based on input queries, which describe the programming task in natural language, or surrounding context, i.e., the code already written by developers.

However, a uniform definition of the current API recommendation task is still absent, leading the task hard to be followed by potential researchers. Some studies~\cite{raychev2014code,bruch2009learning, robbes2010improving,nguyen2016api, kim2021code} regard the task as a code completion problem, and recommend any code tokens including APIs. These studies focus on improving the prediction results of all the tokens instead of only APIs. Some studies~\cite{rahman2016rack, rahman2018nlp2api, gu2016deep, liu2019generating, huang2018api} recommend relative APIs on different levels given natural language queries. Besides, the evaluation results are difficult to be reproduced by future related work. For example, for query-based API recommendation, manual evaluation is generally adopted, so the performances reported by different studies are difficult to be aligned. Comparing with widely-used IDEs or search engines is another commonly adopted yet inconsistent evaluation strategy in previous research. Therefore, to better facilitate future exploration of the API recommendation task, in this paper, we summarize the recent related approaches and build a general benchmark named \bench.

To facilitate the benchmark creation, we group the recent related approaches into two categories according to the task definition: query-based API recommendation and code-based API recommendation.

1) \textbf{query-based API recommendation.} Approaches for query-based API recommendation aim at providing related APIs to developers given a query that describes programming requirements in natural language. The approaches can inform developers which API to use for a programming task.

2) \textbf{code-based API recommendation.} Approaches for code-based API recommendation aim at predicting the next API given the code surrounding the point of prediction. They can directly improve the efficiency of coding.

Besides the unreproducible evaluation, the two groups of studies face their own challenges. 1) For query-based approaches, high-quality queries play a critical role in accurate recommendation. However, there may exist a knowledge gap between developers and API designers in choosing terms for describing queries or APIs. For example, developers who do not know the term ``\textit{heterogeneous list}'' in API documents would use other words such as ``\textit{list with different types of elements}'' in the query. Whether current query reformulation techniques are effective for API recommendation and how effective it is are still remaining unexplored. 2) For code-based approaches, the quality of code before the recommendation point also affects the recommendation performance. Generally, the approaches are evaluated by simulating an actual development, i.e., some parts of a project are removed for imitating a limited context. The APIs to recommend may locate in the front, middle, or back of the code, so exploring the impact of different recommendation points is important for understanding the recommendation capability of existing approaches. Other factors such as whether the APIs are standard or user-defined, lengths of given context, and different domains can also influence the recommendation performance, which have not yet been fully investigated.

To comprehensively understand the above challenges, we evaluate the approaches in \bench from various aspects. \bench is built on Python and Java, and involves two datasets for evaluation, named as \benchq and \benchc for query-based and code-based approaches, respectively.
\benchq contains 6,563 Java queries and 4,309 Python queries obtained from Stack Overflow and API tutorial websites. \benchc contains 1,477 Java projects with 1,229,698 source files and 2,223 Python projects with 414,753 source files obtained from GitHub.
Based on \bench, we study the following research questions:
\begin{itemize}
    \item \textbf{RQ1:} How effective are current query-based and code-based API recommendation approaches?
    \item \textbf{RQ2:} What is the impact of query reformulation techniques on the performance of query-based API recommendation?
    \item \textbf{RQ3:} What is the impact of different data sources on the performance of query-based API recommendations?
    \item \textbf{RQ4:} How well do code-based approaches recommend different kinds of APIs?
    \item \textbf{RQ5:} What is the performance of code-based approaches in handling different
    contexts?
    \item \textbf{RQ6:} How well do code-based approaches perform in cross-domain scenarios?
\end{itemize}

\bench involves the implementation of the related approaches proposed in the recent five years, specifically including five query-based approaches and five code-based approaches. In RQ1, we compare the performances of the approaches in \bench.
To answer RQ2 and RQ3, we apply four popular query reformulation techniques to the queries of \benchq and observe the performance of the query-based approaches given reformulated queries. To answer RQ4 to RQ6, we analyze the APIs in \benchc from different aspects and study the performance of code-based approaches under different experimental settings.

\textbf{Key Findings.} Through the large-scale empirical study, we achieve some findings and summarize the key findings as below.

(1) For query-based API recommendation:
\begin{itemize}
    \item Existing approaches achieve higher recommendation performance on the class level than on the method level. Recommending the exact API methods is still a challenging task.
    \item Query reformulation techniques, including query expansion and query modification, are quite effective in improving the performance of query-based approaches.
    \item  Adding data sources such as Q\&A forums and tutorials that are more similar to real-world queries can significantly improve the performance of current approaches. 
\end{itemize}

(2) For code-based API recommendation:
\begin{itemize}
    \item Recent deep learning models such as Transformers show superior performance on this task. Meanwhile, current IDEs can achieve competitive performance as recent pattern-based and learning-based approaches. They work far away from just recommending APIs based on alphabet orders. 
    \item Current approaches are effective to recommend APIs from standard libraries and popular third-party libraries, but their performance drops a lot when recommending user-defined or project-specific APIs.
    \item Approaches trained on one single domain face the problem of cross-domain adaptation.
    Approaches trained on multiple domains achieve satisfying performance when testing on most single domains, and they even outperform those trained on corresponding single domains.
\end{itemize}

Based on the findings, we conclude some implications and suggestions that would benefit future research. On the one hand, query-based API recommendation approaches should be built along with query reformulation techniques to handle queries with different qualities. We also encourage future work to leverage different data sources and few-shot learning methods to address the low resource challenge in query-based API recommendation. On the other hand, we suggest future code-based API recommendation approaches focus on improving the performance of recommending user-defined APIs as it is the major bottleneck.

\textbf{Contributions.} To sum up, our contribution can be concluded as follows.
\begin{itemize}
    \item To the best of our knowledge, we are the first to systematically study both query-based and code-based API recommendation techniques on two large-scale datasets including Java and Python.
    \item We build an open-sourced benchmark named \bench to fairly evaluate query-based and code-based
    approaches.
    \item We study how different settings can impact the performance of current approaches, including query quality, cross domain adaptation, etc.
    \item We conclude some findings and implications that would be important for future research in API recommendation.
\end{itemize}

The rest of this paper is organized as follows. We present the background and regular API recommendation process in Section~\ref{sec:background}. We describe the details of \bench, current baselines and evaluation metrics in Section~\ref{sec:methodology}. Then we introduce the experiment results and potential findings on query-based and code-based API recommendation in Section~\ref{sec:querybased} and Section~\ref{sec:codebased}, respectively. Based on the findings, we conclude some implications and future directions in Section~\ref{sec:discussion}. Finally, we discuss threats to validity and related work in Section~\ref{sec:threat} and Section~\ref{sec:relatedwork}, respectively.

\begin{figure*}[t]
    \centering
    \includegraphics[width = 0.9\textwidth, ]{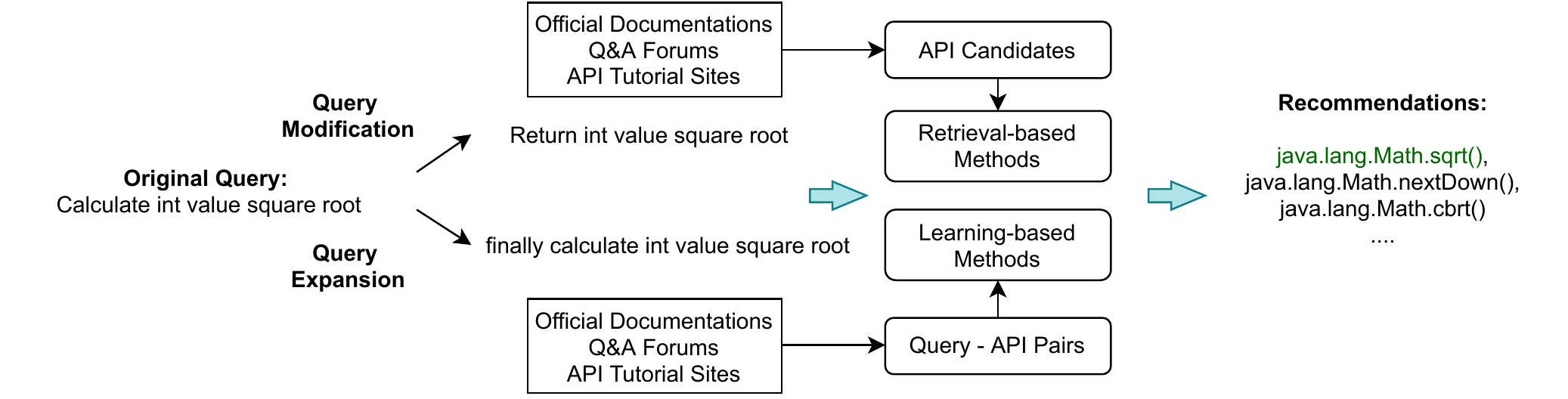}
    \caption{The typical query-based API recommendation framework.}
    \label{fig:querymodel}
\end{figure*}

\begin{figure*}[t]
     \centering
     \includegraphics[width = 1.0\textwidth]{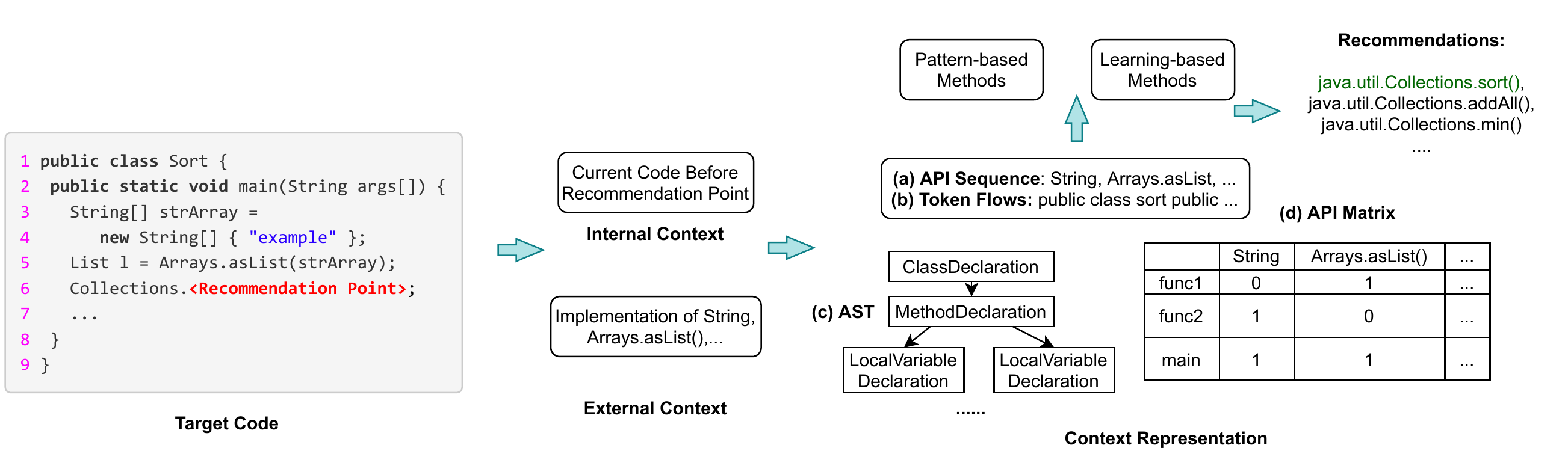}
     \caption{The typical code-based API recommendation framework.}
     \label{fig:codemodel}
\end{figure*}

\section{Background}\label{sec:background}
In this section, we summarize the query-based approaches and code-based approaches, respectively.

\subsection{Query-Based API Recommendation Methods}\label{sec:querybackground}
We describe the typical query-based API recommendation process in figure~\ref{fig:querymodel}. Given a query ``\textit{Calculate int value square root}'', query reformulation techniques first modify the query as ``\textit{return int value square root}'' or expand it as ``\textit{finally calculate int value square root}''. A knowledge base built upon available data sources is also prepared for API candidate selection. Based on the knowledge base, retrieval-based methods or learning-based methods recommend the APIs relevant to the queries.

\subsubsection{Query Reformulation Techniques}
Input queries can be short in length or vague in semantics. Besides, there may exist a knowledge gap between developers and search engines in query description. For rendering search engines better understand the query semantics, query reformulation is a common pre-processing method. In general, there are two major types of query reformulation approaches:

1) query expansion, which adds extra information to the original queries; 

2) query modification, which modifies, replaces or deletes some words in the original queries.

\textbf{Query expansion.} Query expansion aims at identifying important words that are missing in the input queries. The topic is originally stemmed from the field of natural language processing (NLP). For example, the work~\cite{Lu2015queryexpansion} utilizes word embeddings to map words in the vector space and finds similar words to enrich the queries. For the API recommendation task, since APIs are encapsulated and organized according to classes and modules, class names and module names are important hints for recommendation. Rahman \etal~\cite{rahman2016rack,rahman2018nlp2api} propose to use keyword-API class co-occurrence frequencies and keyword-keyword co-occurrence frequencies to build the relationship between words and API classes, and add the suggested API class for query expansion.

\textbf{Query modification.} Query modification aims at mitigating both the lexical gap and knowledge gap between the user queries and descriptions in knowledge base. The lexical gap ,such as mis-spelling, can be easily addressed by spelling correction and synonym search, etc.
Recent work focuses on how to mitigate the knowledge gap by replacing inappropriate words in queries. Mohammad \etal~\cite{mohammad2017improved} extract important tokens in code, and Sirres \etal~\cite{sirres2018augmenting} leverage discussions and code from Stack Overflow posts to build a knowledge base. Cao \etal~\cite{cao2021automated} collect query reformulation history from Stack Overflow and propose a Transformer-based approach to learn how developers change their queries when search engines do not return desired results.

\subsubsection{Recommendation with Knowledge Base}

\textbf{Knowledge base.} API recommendation approaches generally require a knowledge base that contains all the existing APIs as the search space. There are three primary sources for the knowledge base creation, including: 
1) official documentations which contain comprehensive descriptions about the API functionality and structure.
2) Q\&A forums, which provide the purposes of APIs and different API usage patterns. Many studies~\cite{huang2018api,rahman2018effective} leverage the Q\&A pairs from Stack Overflow to select API candidates. 
3) Wiki sites, which describe concepts that link different APIs. For example, Liu \etal~\cite{liu2019generating} utilizes API concepts from Wikipedia to help build API knowledge graphs.

\textbf{Retrieval-based methods.} Retrieval-based methods retrieve API candidates from the knowledge base and then rank the candidate APIs by calculating the similarities between queries and APIs. For example, Rahman \etal~\cite{rahman2016rack,rahman2018nlp2api} utilize the keyword-API occurrence frequencies and API-API occurrence frequencies to find the most relevant APIs. Huang \etal ~\cite{huang2018api} first identify the similar posts from Stack Overflow by computing query-documentation similarities and choose the APIs mentioned in posts as candidates. Liu \etal~\cite{liu2019generating} build an API knowledge graph to represent relationships between APIs and then calculate the similarities between queries and certain parts of API knowledge graph to rank the APIs.

\textbf{Learning-based methods.} Another type of method is to automatically learn the relationships between queries and APIs based on deep learning techniques. The knowledge base provides query-API pairs as the ground truth. For example, Gu \etal~\cite{gu2016deep} formulate the task as a translation problem in which a model is built to translate word sequences into API sequences. They propose an RNN model with a encoder-decoder structure to implement the translation.

\subsection{Code-Based API Recommendation Methods}
We describe the workflow of code-based API recommendation in Figure~\ref{fig:codemodel}. Given a target code, context representation is an essential step. Based on the extracted context, pattern-based methods or learning-based methods are adopted by previous studies to recommend the next API.

\subsubsection{Context for the Target Code}
Most code-based API recommendation methods regard the code before the recommendation point as the context. We name such context as \textit{internal context} since it only considers code in the current source code or current function body. For example, Line 1 $\sim$ 6 of the target code in Figure~\ref{fig:codemodel} belongs to internal context. Xie \etal~\cite{xie2019hirec} find that replacing external APIs in code (such as $Arrays.asList()$ in Figure~\ref{fig:codemodel}) with their implementations can help the identification of common usage patterns. They propose to build a hierarchical context by integrating the implementation out of the current source file. We name the implementation of external APIs as \textit{external context}.

\subsubsection{Context Representation}
We divide the context representation methods into two types, i.e., pattern-based representation and learning-based representation. \textbf{Pattern-based representations}~\cite{wen2021siri, xie2019hirec, nguyen2012graph, nguyen2019focus, d2016collective} do not consider all the code tokens. Instead, they only identify APIs to build API usage sequences, as shown in fig.~\ref{fig:codemodel} (a), API matrix, as shown in fig.~\ref{fig:codemodel} (d), or API dependency graphs to represent the current context. \textbf{Learning-based representations}~\cite{hindle2016naturalness, tu2014localness, raychev2014code,he2021pyart,kim2021code} usually represent the context with token flows, as illustrated in Figure~\ref{fig:codemodel} (b), or other syntax structures such as Abstract Syntax Trees (ASTs), as illustrated in Figure~\ref{fig:codemodel} (c).

\subsubsection{Recommendation Based on Context}
\textbf{Pattern-based methods.} API recommendation is inherently a recommendation task, so some studies~\cite{d2016collective,nguyen2019focus} follow the collaborative filtering (\textit{user-item}) methodology of traditional recommendation systems~\cite{Ricci2011}. As shown in Figure.~\ref{fig:codemodel} (d), they regard the internal context as the \textit{users} and APIs as the \textit{items}. They then calculate the similarities between different \textit{users} to find the most similar API for recommendation. However, the methods do not consider the relationships between APIs. More recent work~\cite{wen2021siri,xie2019hirec} build API dependency graphs or mines association rules to capture API usage patterns. 

\begin{table*}[th]
    \centering
    \caption{Statistics of Benchmark \benchc. The data includes both the training set and testing set.}
    \scalebox{0.93}{
    \begin{tabular}{c|rrrrrrrrrrr}
    \toprule
         \multirow{2}*{\textbf{PL}} &  \multirow{2}*{\textbf{Domain}} & \multirow{2}*{\textbf{\#Projects}} & \multirow{2}*{\textbf{\#Files}} & \multirow{2}*{\textbf{\tabincell{c}{LOC \\ (per func)}}} & \multirow{2}*{\textbf{\tabincell{c}{\#API \\ (per func)}}} & \multicolumn{3}{c}{\textbf{Total number of APIs (only testset)}}  & \multirow{2}*{\textbf{\tabincell{c}{LOC Threshold\\ of Short Func}}} & \multirow{2}*{\textbf{\tabincell{c}{LOC Threshold \\ of Long Func }}}\\
         & & & & & & \textbf{Standard} & \textbf{User-defined} & \textbf{Popular} & & \\
         \midrule
         \multirow{6}*{\textbf{Python}} & General & 899 & 230,064 & 15.24 & 5.55 & 1,363,240 & 1,747,878 & 54,244 & 8.875 & 54.875 \\
         & ML & 323 & 46,556 & 13.89 & 6.08 & 629,437 & 339,821 & 125,377 & 12.65 & 46.05 \\
         & Security & 126 & 15,785 & 18.98 & 6.72 & 111,393 & 64,809 & 3,613 & 6 & 86.5 \\
         & Web & 568 & 82,771 & 14.14 & 5.05 & 369,114 & 241,602 & 11,832 & 7.35 & 51.625 \\
         & DL & 307 & 39,577 & 14.58 & 6.25 & 413,295 & 220,228 & 76,654 & 11.675 & 52.525 \\
         \midrule
         \multirow{5}*{\textbf{Java}} & General & 935 & 1,056,790 & 11.16 & 4.06 & 5,164,481 & 3,808,124 & 36,178 & 6.26 & 19.2 \\
         & Android & 377 & 87,468 & 8.24 & 2.91 & 517,461 & 267,141 & 75,069 & 7.28 & 16.8 \\
         & ML & 52 & 41,377 & 12.82 & 4.77 & 194,013 & 136,963 & 0 & 7.52 & 19.74 \\
         & Testing & 55 & 23,618 & 9.93 & 3.98 & 105,577 & 55,241 & 22 & 6.44 & 15.68 \\
         & Security & 58 & 20,445 & 12.35 & 5.32 & 125,558 & 74,471 & 1,243 & 6.88 & 20.78 \\
    \bottomrule
    \end{tabular}}
    \label{tab:benchc}
\end{table*}

\textbf{Learning-based methods.} Hindle \etal~\cite{hindle2016naturalness} discover the naturalness of software, rendering it possible to deploy machine learning or deep learning
methods on code. Different from pattern-based methods that consider the relationships between API occurrences, learning-based methods regard API as a single code token, and reformulate the code-based API recommendation problem into a \textit{next token prediction} problem. Many statistical language models~\cite{raychev2014code, nguyen2015graph, tu2014localness, Raychev2016prob} are proposed to predict the next code token. Besides using the token sequences, more recent work~\cite{he2021pyart, kim2021code} try to leverage syntax and data flow information for more accurate prediction.

\begin{table}[htb]
    \centering
    \caption{Statistics of \benchq. Ori. represent the original queries, Exp. represent the expanded queries produced by query expansion techniques, Mod. represent the modified queries produced by query modification techniques.}
    \begin{tabular}{c|rrr|rrr}
    \toprule
         \multirow{2}*{\textbf{PL}} & \multicolumn{3}{c}{\textbf{Stack Overflow}} & \multicolumn{3}{c}{\textbf{Tutorial Websites}} \\
         & Ori. & Exp. & Mod. & Ori. & Exp. & Mod. \\
         \midrule
        \textbf{Python} &  1,925 & 78,157 & 100,100 & 2,384 & 95,360 & 123,968 \\
        \textbf{Java} & 1,320 & 80,343 & 68,640 & 5,243 & 319, 783 & 272,636 \\
        \bottomrule
    \end{tabular}
    \label{tab:benchq}
\end{table}

\section{Methodology}\label{sec:methodology}
In this section, we introduce the scope of the studied APIs, the preparation of benchmark datasets, and implementation details.
 
\subsection{Scope of APIs}\label{sec:api}
To fairly compare the current API recommendation approaches, a benchmark dataset should be prepared, during which the scope of studied APIs firstly needs to be defined. In this work, we focus our evaluation on two popular programming languages, i.e., Python and Java.

For facilitating the analysis of the challenges in API recommendation, we divide all APIs into \textit{standard APIs}, \textit{user-defined APIs}, and \textit{popular third-party APIs}. The \textit{standard APIs} refer to the APIs that are clearly defined and built-in in corresponding programming languages while the \textit{user-defined APIs} are defined and used in projects along with \textit{popular third-party APIs}. Note that we evaluate query-based API recommendation methods only on the \textit{standard APIs} since \textit{user-defined APIs} are generally not associated with detailed descriptions or extensive discussions for facilitating the recommendation. We evaluate code-based API recommendation methods on all three kinds of APIs. The details of each kind for different programming languages are depicted below.



\textbf{(1) Standard Java APIs.} We choose the version Java 8 for our analysis since it is the most widely-used version in current projects according to the 2020 JVM Ecosystem Report~\cite{javasurvey}. We collect 34,072 APIs from the Java documentation~\cite{javadoc} as \textit{standard APIs}.


\textbf{(2) Popular Java third-party APIs.} We choose APIs from the Android library~\cite{androiddoc} since Android is one of the most popular applications of Java programs. We collect 11,802 APIs from the official documentation of Android in total and regard them as the popular third-party APIs.

\textbf{(3) Standard Python APIs.} As Python Software Foundation has stopped the support for Python 2, currently only 6\% of developers are still using Python 2, according to the development survey conducted by Jetbrains~\cite{jetbrainssurvey}. Considering that APIs of different versions above 3.0 are similar, we choose the newest version 3.9 to ensure the compatibility, and collect 5,241 APIs from Python standard library~\cite{pythonlib} as the \textit{standard APIs}. 

\textbf{(4) Popular Python third-party APIs.} Python is well extended by a lot of third-party modules. We choose five widely-used modules with sufficient documentations, including flask~\cite{flaskdoc}, django~\cite{djangodoc}, matplotlib~\cite{matplotlibdoc}, pandas~\cite{pandasdoc} and numpy~\cite{numpydoc}. We collect 215, 700, 4,089, 3,296 and 3,683 APIs from them, respectively.

\textbf{(5) User-defined APIs.} For code-based API recommendation, we regard all the functions defined in current projects as \textit{user-defined APIs}. We do not explicitly collect them as a fixed set because they vary across projects. By inspecting the implementations, we can always identify the user-defined APIs.

\subsection{Benchmark Dataset}
In this section, we describe how we build the benchmark datasets \benchq and \benchc.

\subsubsection{Creation of \benchq}
We build the benchmark dataset \benchq by mining Stack Overflow and tutorial websites. Note that we find that currently there is no query-based API recommendation approach specially designed for Python programs, but we still collect the query benchmark for it to facilitate further research.

\textbf{Mining Stack Overflow.} As one of the most popular Q\&A forums for developers, Stack overflow contains much discussion about the usage of APIs. Stack Overflow is the primary source for building \benchq. We first download all posts from Aug 2008 to Feb 2021 on Stack Overflow (SO) via Stack Exchange Data Dump~\cite{sodump}. Each post is associated with a tag about the related programming language. We filter out the posts not tagged as Java or Python, resulting in 1,756,183 Java posts and 1,661,383 Python posts. We further filter out the posts based on the following rules:

\begin{itemize}
    \item To increase the quality of the posts, we remove the posts that do not have endorsed answers.
    \item We remove the posts whose titles contain keywords such as ``error" and ``why"  that are rarely used by programmers for asking about API recommendation.
    \item We remove the posts that do not contain the HTML tag $<$code$>$, because we cannot extract any API from them.
    \item We remove the posts that contain code snippets longer than two lines, since large code snippets in a post usually indicate that the query should be handled by a series of operations, not a single API sequence.
    \item We check the code in each post and remove the posts that do not contain any APIs involved in the paper, as described in Sec.~\ref{sec:api}.
\end{itemize}

After the rule-based filtering, we obtain 156,493 Python posts and 148,938 Java posts that contain descriptions about APIs. However, some of the posts are not directly related to API recommendation. For example, some posts only ask about comparing two similar APIs. The unrelated posts are hard to be automatically identified by rules. To ensure the relatedness of the posts in our benchmark dataset, we invite 16 participants with an average of 3-year development experience in Python or Java for manually checking. For each post, two of the participants are involved to check the following aspects:


1) whether the query asks about API recommendation;

2) whether the standard APIs recognized by the previous rules are intact, i.e., including the whole class and method names.

3) whether the APIs in answers exactly address the query.

\noindent If two participants provide the same answers for one post and also one of the above three aspects is not satisfied, we directly remove the post. If the two participants do not reach an agreement, the post will be forwarded to one of the authors to make a final decision. 

We manually check 13,775 posts, in which 1,262 posts do not reach an agreement by the annotators and need further check by one of the authors.
We use the commonly-used Fleiss Kappa score~\cite{Fleiss1971kappa} to measure the agreement degree between the two annotators and the value is 0.77. The result indicates a high agreement between them. Based on the manual check, 3,245 of the 13,775 labeled posts are remained, including 1,925 Python queries and 1,320 Java queries, comprising the first part of our benchmark \benchq, as shown in the second column of Table~\ref{tab:benchq}.


\textbf{Mining tutorial websites.} API tutorial websites are the second major source of query-API pairs. We choose three popular API tutorial websites GeeksforGeeks~\cite{g4g}, Java2s~\cite{java2s} and Kode Java~\cite{kodejava} to establish \benchq. Different from Stack Overflow that contains discussion on various topics, API tutorial websites focus on providing examples of how to use APIs. Therefore, manually annotating the relatedness of each query to API recommendation is not necessary. We adopt similar rules as mining Stack Overflow to filter out those without code snippets or associated with large code snippets. We finally collect 5,243 Java queries and 2,384 Python queries, which comprise the second part of our benchmark \benchq, as shown in the fifth column of Table~\ref{tab:benchq}. 


\begin{figure}[tb]
    \centering
    \begin{minipage}[t]{0.49\linewidth}
    \centering
    \includegraphics[width = 1.0\linewidth]{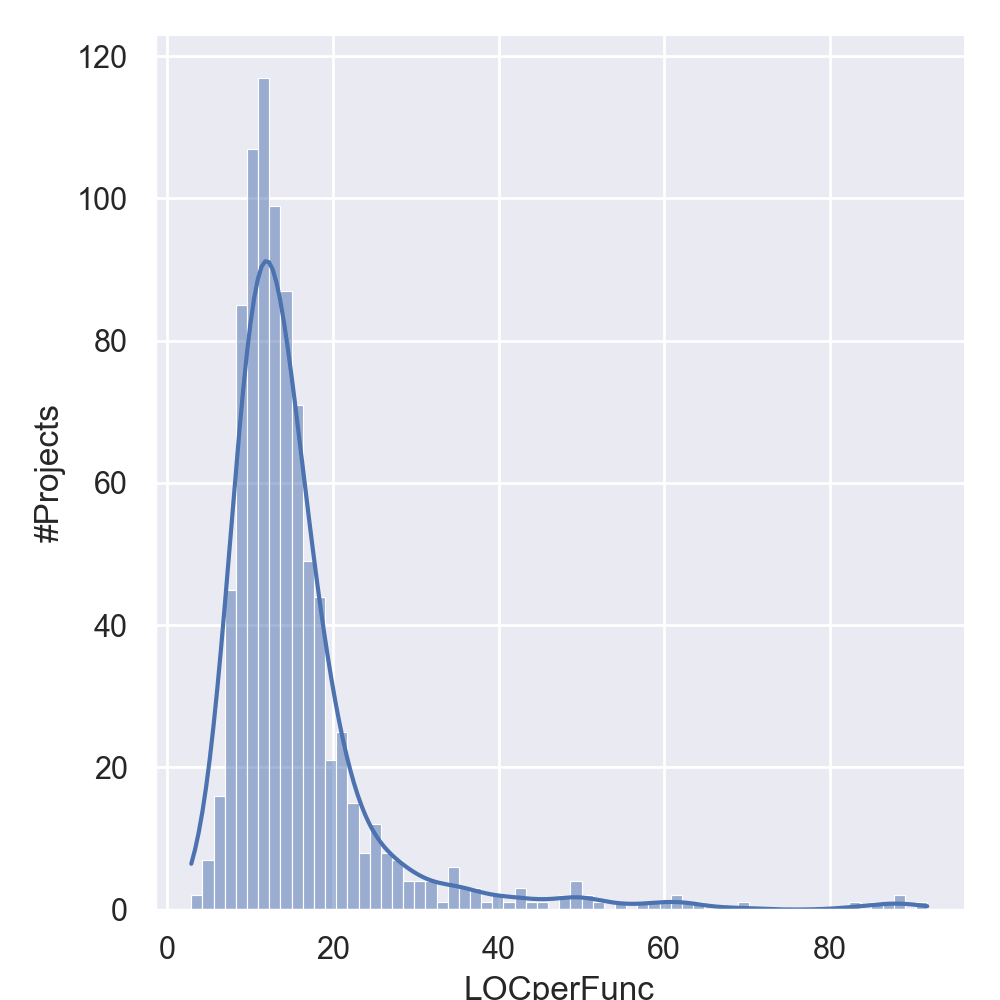}
    \end{minipage}
    \begin{minipage}[t]{0.49\linewidth}
    \centering
    \includegraphics[width = 1.0\linewidth]{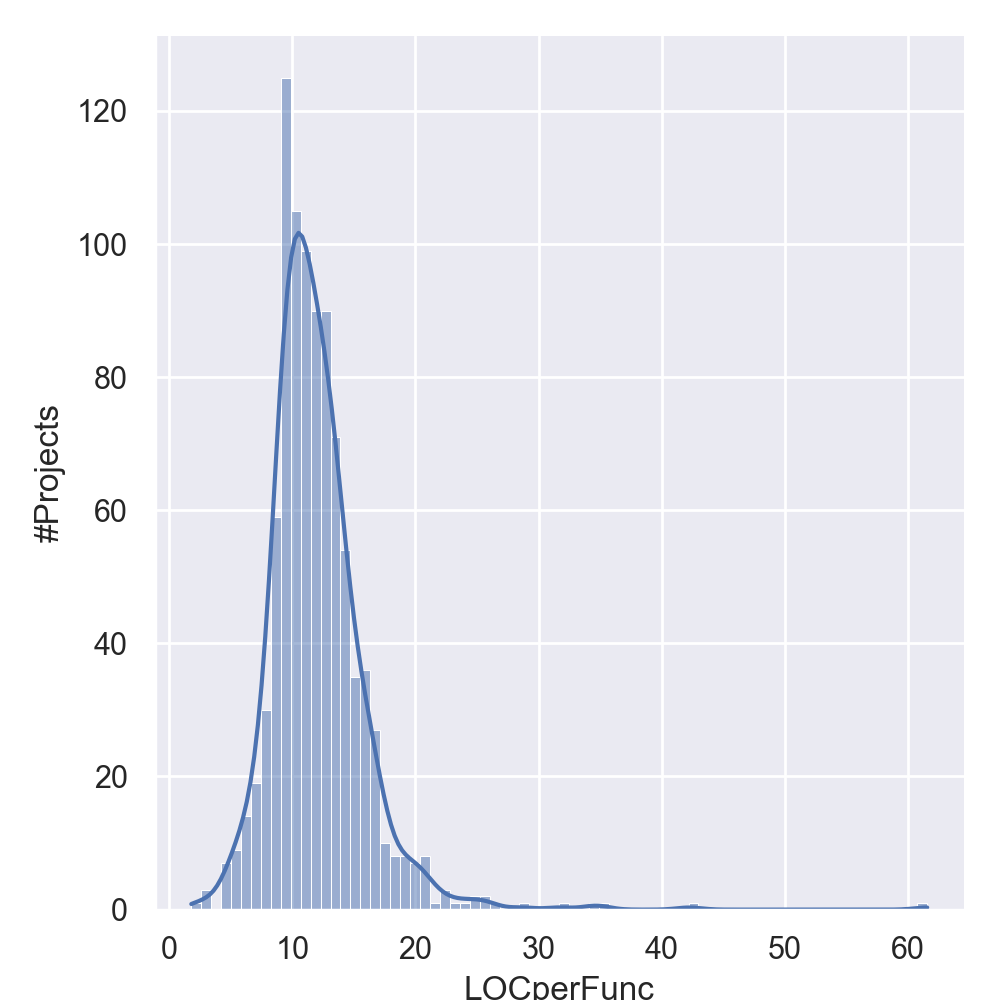}
    \end{minipage}
    \caption{Distribution of code lines per function at \textit{general} domain (Left: Python, Right: Java).}
    \label{fig:lengthdist}
\end{figure}

\begin{table*}[th]
    \centering
    \caption{The query reformulation techniques and query-based API recommendation approaches involved in the paper. For tools, the years they last updated are listed. The column name ``PL'' indicates the applicable programming language.
    }
    \begin{tabular}{ccp{8.5cm}ccc}
        \toprule
        \multirow{2}*{\textbf{Approach}} & \multirow{2}*{\tabincell{c}{\textbf{Category/} \\ \textbf{Data Source}}} & \multirow{2}*{\textbf{Description}} &  \multirow{2}*{\textbf{\tabincell{c}{PL}}} & \multirow{2}*{\textbf{Venue}} & \multirow{2}*{\textbf{Year}} \\
        & & & & \\
        \midrule
        \rowcolor{Gray}
        \multicolumn{6}{c}{\textbf{Query Reformulation}}\\
        \midrule
         \multirow{2}*{\tabincell{c}{Google \\ Prediction Service}} & \multirow{2}*{\tabincell{c}{Query expansion, \\ modification}}  & Google Prediction Service~\cite{gps} is a common query reformulation technique used in Google Search. & \multirow{2}*{Any} & \multirow{2}*{-} & \multirow{2}*{2021}\\
        \midrule
         \multirow{3}*{\tabincell{c}{NLPAUG}} & \multirow{3}*{\tabincell{c}{Query expansion, \\ modification}}  & NLPAUG~\cite{ma2019nlpaug} is a library that integrates data augmentation techniques in the NLP field. It provides character-level, word-level and sentence-level reformulations to a certain sentence based on DL models or random methods. & \multirow{3}*{Any} & \multirow{3}*{-} & \multirow{3}*{2021} \\
         \midrule
         \multirow{3}*{\tabincell{c}{SEQUER}} & \multirow{3}*{\tabincell{c}{Query expansion, \\ modification}} & SEQUER~\cite{cao2021automated} is a Transformer trained for query reformulation based on the query logs from Stack Overflow. We access the tool via the demo link~\cite{sequer} provided by the authors. & \multirow{3}*{Any} & \multirow{3}*{ICSE} & \multirow{3}*{2021} \\
         \midrule
         \multirow{3}*{\tabincell{c}{NLP2API}} & \multirow{3}*{\tabincell{c}{Query expansion}} & NLP2API~\cite{rahman2018nlp2api} is a query expansion approach that adds predicted API class into the original query. We re-implement it based on the replication package~\cite{nlp2api} released by the authors. & \multirow{3}*{Java} & \multirow{3}*{ICSME} & \multirow{3}*{2018} \\
         \midrule
         \rowcolor{Gray}
        \multicolumn{6}{c}{\textbf{Query-Based API Recommendation}}\\
       \midrule
       \multirow{3}*{\tabincell{c}{RACK}} & \multirow{3}*{\tabincell{c}{Official documentation, \\ Stack Overflow}}  & RACK~\cite{rahman2016rack} is a class-level recommendation approach that leverages keyword and API co-occurrence information to make predictions. We re-implement the approach based on the released replication package~\cite{rack}.& \multirow{3}*{Java} & \multirow{3}*{ICSE} & \multirow{3}*{2016} \\
         \midrule
         \multirow{3}{*}{\tabincell{c}{KG-APISumm}} & \multirow{3}{*}{\tabincell{c}{Official documentation, \\ Wikipedia}} & KG-APISumm~\cite{liu2019generating} is a class-level recommendation approach that builds an API knowledge graph to provide API class recommendations with detailed summaries.& \multirow{3}*{Java} & \multirow{3}*{FSE} & \multirow{3}*{2019} \\
         \midrule
         \multirow{4}{*}{\tabincell{c}{Naive Baseline}} & \multirow{4}{*}{\tabincell{c}{Official documentation}}  & We build
         an API-level naive baseline that only calculates the similarity score between API descriptions provided by
         official documentation and queries based on BERTOverflow~\cite{bertovereflow} to recommend API classes. It indicates the basic performance of similarity-based methods. & \multirow{4}*{Any} & \multirow{4}*{-} & \multirow{4}*{2021} \\
         \midrule
         \multirow{4}{*}{\tabincell{c}{DeepAPI}} & \multirow{4}{*}{\tabincell{c}{Official documentation}} & DeepAPI~\cite{gu2016deep} is an API-level recommendation approach that uses an encoder-decoder model to convert words into APIs. We train this model using the training set released by its authors for 275,000 iterations and evaluate its performance on \benchq. & \multirow{4}*{Java} &\multirow{4}*{FSE} & \multirow{4}*{2016} \\
         \midrule
         \multirow{3}*{Lucene} & \multirow{3}*{Official documentation} & Lucene~\cite{lucene} is a popular java library providing powerful indexing and search features. Given a query, it can be used as a search engine to search related APIs in the official documentation. & \multirow{3}*{Any} & \multirow{3}*{-} & \multirow{3}*{2021} \\
         \midrule
         \multirow{5}{*}{\tabincell{c}{BIKER}} & \multirow{5}{*}{\tabincell{c}{Official documentation, \\ Stack Overflow}} & BIKER~\cite{huang2018api} is a method-level recommendation approach that leverages both Stack Overflow posts and API documentations to recommend APIs for a programming task. It fetches the top-k similar Stack Overflow posts to narrow down the scope of API candidates. We re-implement the approach based on the released replication package~\cite{biker}. & \multirow{5}*{Java} & \multirow{5}*{ASE} & \multirow{5}*{2018} \\
        \bottomrule
    \end{tabular}
    \label{tab:querybaselines}
\end{table*}
\begin{table*}[th]
    \centering
    \caption{The code based API recommendation baselines included in this empirical study. For tools we list the year of its most recent update time. The column name ``PL'' indicates the applicable programming language.}
    \begin{tabular}{ccp{9cm}ccc}
        \toprule
        \multirow{2}*{\textbf{Approach}} & \multirow{2}*{\textbf{Representation}} & \multirow{2}*{\textbf{Description}} &  \multirow{2}*{\textbf{\tabincell{c}{PL}}} & \multirow{2}*{\textbf{Venue}} & \multirow{2}*{\textbf{Year}} \\
        & & & & \\
        \midrule
        \rowcolor{Gray}
        \multicolumn{6}{c}{\textbf{Practical IDE}}\\
        \midrule
        \multirow{3}*{\tabincell{c}{PyCharm}} & \multirow{3}*{\tabincell{c}{Code tokens}} & PyCharm~\cite{pycharm} is a widely-used Python IDE released by JetBrains. We use its
        basic code completion feature~\cite{pycharmcodecompl} to recommend APIs given an incomplete piece of code. & \multirow{3}*{Python} & \multirow{3}*{-} & \multirow{3}*{2021} \\
         \midrule
         \multirow{4}*{\tabincell{c}{Visual Studio Code}} & \multirow{4}*{\tabincell{c}{Code tokens}} & Visual Studio Code~\cite{vscode} is a widely-used code editor released by Microsoft. We use its popular
         \textit{Python} extension (ID: ms-python.python) that includes IntelliSense (Pylance) to recommend APIs given an incomplete piece of
         code. & \multirow{4}*{Python}& \multirow{4}*{-} & \multirow{4}*{2021} \\
         \midrule
         \multirow{3}*{\tabincell{c}{Eclipse}} & \multirow{3}*{\tabincell{c}{Code tokens}} & Eclipse~\cite{eclipse} is a widely-used Java IDE released by Eclipse Foundation. We use its default code completion feature to recommend APIs given an incomplete piece of
         code. & \multirow{3}*{Java}& \multirow{3}*{-} & \multirow{3}*{2021} \\
         \midrule
         \multirow{3}*{\tabincell{c}{IntelliJ IDEA}} & \multirow{3}*{\tabincell{c}{Code tokens}} & IntelliJ IDEA~\cite{IntelliJ} is a widely-used Java IDE released by JetBrains. We use the default code completion feature to recommend APIs given an incomplete piece of Java code. & \multirow{3}*{Java}& \multirow{3}*{-} & \multirow{3}*{2021}\\
         \midrule
         \rowcolor{Gray}
        \multicolumn{6}{c}{\textbf{Approach in Academia}}\\
         \midrule
         \multirow{4}*{\tabincell{c}{TravTrans}} & \multirow{4}*{\tabincell{c}{AST}} & TravTrans~\cite{kim2021code} encodes AST  node sequences generated by pre-order traversal into Transformer and predicts the masked AST nodes that correspond to some certain tokens in the source code. We re-implement it based on the replication package~\cite{facebooktransformer} provided by the authors. & \multirow{4}*{Python}& \multirow{4}*{ICSE}  & \multirow{4}*{2021} \\
         \midrule
         \multirow{4}*{\tabincell{c}{PyART}} & \multirow{4}*{\tabincell{c}{Token flow \\ Data flow}} & PyART~\cite{he2021pyart} leverages a predictive model trained on data flow information, token similarity and token co-occurrence based on the Random Forest algorithm to recommend APIs. We re-implement it based on the replication package~\cite{pyart} provided by the authors. & \multirow{4}*{Python}& \multirow{4}*{ICSE} & \multirow{4}*{2021} \\
         \midrule
         \multirow{4}*{\tabincell{c}{Deep3}} & \multirow{4}*{\tabincell{c}{AST, DSL}} & Deep3~\cite{Raychev2016prob} is a general approach for building probabilistic models of code based on learning decision trees. The decision trees are represented as program in a domain specific language (DSL) called TGEN which is agnostic to programming languages. We re-implement it based on the replication package~\cite{deep3} provided by the authors. & \multirow{4}*{Python}& \multirow{4}*{ICML} & \multirow{4}*{2016} \\
         \midrule
         \multirow{5}*{\tabincell{c}{FOCUS}} & \multirow{5}*{\tabincell{c}{API Matrix}} & FOCUS~\cite{nguyen2019focus} leverages the idea of traditional recommendation systems and uses a 3D matrix to represent the relationship between different projects, functions and APIs. It recommends APIs and usage patterns from the most similar projects. We re-implement it based on the released replication package~\cite{focus}. 
         & \multirow{5}*{Java}& \multirow{5}*{ICSE} & \multirow{5}*{2019} \\
         \midrule
         \multirow{3}*{\tabincell{c}{PAM}} & \multirow{3}*{\tabincell{c}{API sequence}} & PAM~\cite{fowkes2016parameter} builds a probabilistic model to mine the most ``interesting''  API patterns in certain projects. We re-implement it based on the released replication package~\cite{pam}.
         & \multirow{3}*{Java} & \multirow{3}*{FSE} & \multirow{3}*{2016} \\
         \midrule
         \multirow{4}*{\tabincell{c}{PAM-MAX}} & \multirow{4}*{\tabincell{c}{API sequence}} & As PAM is originally a context-insensitive API recommendation approach, we build PAM-MAX by selecting the most correct API recommendation among all API patterns mined in the training set. It indicates the theoretical maximum performance a context-insensitive approach can reach when it can always find the most related projects to mine API patterns.
         & \multirow{4}*{Java} & \multirow{4}*{FSE} & \multirow{4}*{2016} \\
         \bottomrule
    \end{tabular}
    \label{tab:codebaselines}
\end{table*}

\subsubsection{Creation of \benchc}

We create the benchmark dataset \benchc by mining GitHub. GitHub~\cite{github} is one of the most popular websites for sharing code and includes large numbers of code repositories on different topics and programming languages. 

In order to explore the performance of API recommendation under different domains, we first determine the domains for analysis. According to the JetBrains' developer survey~\cite{jetbrainssurvey} and topic labels provided by GitHub, we choose four popular domains for Python and Java, respectively, as shown in Table~\ref{tab:benchc}. For Python, we consider the domains ``ML'', ``Security'', ``Web'', and ``DL''; while for Java, we involve domains ``Android'', ``ML'', ``Testing'', and ``Security''. For each domain, we collect 500 repositories with the most stars and 500 repositories with the most forks on GitHub\footnote{The collection was conducted during April 2021.}. Besides the specific domains, we also build a ``General'' domain which only considers the popularity of repositories. For the ``General'' domain, we collect 1,000 repositories with the most stars and 1,000 repositories with the most forks on GitHub regardless of the topics.

Not all the collected repositories are applicable for code-based API recommendation. Some popular repositories do not contain enough code, e.g., only including documentations. To remove such repositories, we use \textit{cloc}~\cite{cloc} to scan the code in each repository and filter out the repositories that 1) have fewer than 10 files or 2) have fewer than 1000 lines of code or 3) have code in Python or Java but with the ratio less than 10\%. The number of projects, number of files, and average number of code lines for each domain of \benchc are shown in Table~\ref{tab:benchc}. 


As most approaches~\cite{kim2021code, raychev2014code, he2021pyart,nguyen2019focus,fowkes2016pam} for code-based API recommendation require a training set to learn the API patterns or train the models, we split \benchc into a training set and a test set with a ratio of 80\% and 20\%, respectively. Note that we do not split a project both into the training set and test set, but put all the files of the same project into either the training set or test set, because Alon \etal~\cite{alon2019code2seq} and LeClair \etal~\cite{alexander2019nlsummaries} find that code in the same project usually share the same variable names and code patterns, and splitting without considering project can cause data leakage. For the approaches requiring a validation set, we prepare it from the training set.

In order to study the impact of different recommendation points and different lengths of functions on the performance of current approaches, we analyze the average length of functions in each repository. By using Kernel Density Estimation (KDE), we observe that the length distributions of functions in different domains are similar. The distributions of the ``General'' domain are depicted in Figure~\ref{fig:lengthdist}. From the figure we observe that most Python functions contain 5 $\sim$ 30 lines of code (LOC) and most Java functions contain 5 $\sim$ 20 lines of code. However, there still exist a few extremely short or long functions. These functions are likely to impact the performance of code-based approaches with extremely long or short contexts. To derive the functions of appropriate lengths for study, we regard the shortest 5\% of the functions as extremely short functions, the longest 5\% of the functions as extremely long functions, and the middle 90\% of the functions as functions with moderate lengths. We then further study the performance of code-based approaches on them in sec.~\ref{sec:rq5}. The thresholds for distinguishing short and long functions regarding the Line of code are illustrated in the last two columns of Table~\ref{tab:benchc}.

In order to study the performance of current approaches on different kinds of APIs, we convert source files of each repository into ASTs and extract all the function calls in them. We label a function call as a \textit{standard} API or \textit{popular} third-party API if it matches one of the APIs collected in Sec.~\ref{sec:api}. We label a function call as a \textit{user-defined} API if its implementation can be found in the current repository via import analysis. The average number of APIs per function, number of \textit{standard} APIs and number of \textit{user-defined} APIs are shown in column 6 $\sim$ 8 of Table~\ref{tab:benchc}.

\subsection{Implementation Details}

In this section, we describe the details of each approach involved in the benchmark and the metrics for evaluation.


\textbf{Query reformulation techniques.} We choose four popular query reformulation techniques, including Google Prediction Service~\cite{gps}, NLPAUG~\cite{ma2019nlpaug}, SEQUER~\cite{cao2021automated}, and NLP2API~\cite{rahman2018nlp2api}. The detailed description of each technique is illustrated in Table~\ref{tab:querybaselines}. Google prediction service is included as one of the most effective approaches in practice, while SEQUER~\cite{sequer} is the state-of-the-art approach. NLPAUG~\cite{ma2019nlpaug} is considered since it is widely used for query reformulation in many NLP studies~\cite{raghu2020survey, xue2021robustness, jing2021identifying, newman2021padapters}. We also include NLP2API~\cite{nlp2api} since it differs from major reformulation methods by first predicting the API class related to the query and then adding the predicted API class into the query.


\textbf{Query-based API recommendation approaches.} We choose five query-based API recommendation approaches published by recent top conferences, including KG-APISumm~\cite{liu2019generating}, BIKER~\cite{huang2018api}, RACK~\cite{rahman2016rack}, and DeepAPI~\cite{gu2016deep}, along with a popular search library Lucene~\cite{lucene}. The detailed description of each baseline is shown in Table~\ref{tab:querybaselines}. We reproduce the five approaches based on the replication packages released by the authors. Besides, we build a naive baseline that recommends APIs by computing the similarities between queries and API descriptions based on BERTOverflow~\cite{bertovereflow}. The native baseline serves as an indicator of the basic performance of similarity-based models. We also notice that different sources are adopted by the approaches for creating the knowledge base. For example, the naive baseline and DeepAPI only consider official documentation, while BIKER and RACK also involve the Q\&A forum -- Stack Overflow. We list the knowledge source of each approach in Table~\ref{tab:querybaselines}. During implementation, we do not align the sources of the approaches, since the sources are claimed as contributions in the original papers. Instead, we design a separate RQ to study the impact of knowledge sources on the performance of API recommendations.



During studying the impact of query reformulation on the recommendation performance, we implement all the four query reformulation techniques for each of the six API recommendation baselines because all the baselines do not integrate query reformulation techniques in the original papers.

\textbf{Code-based API recommendation approaches.} We choose four IDEs and five approaches published on recent top conferences as our code-based API recommendation baselines. A detailed description of each baseline is shown in Table~\ref{tab:codebaselines}. For the IDEs and some of the approaches such as TravTrans~\cite{kim2021code} and Deep3~\cite{Raychev2016prob}, they can predict any code tokens besides API tokens. In the paper, we focus on evaluating their performance in recommending APIs, following prior research~\cite{kim2021code, raychev2014code, he2021pyart,nguyen2019focus,fowkes2016pam} we use the training set of \benchc to train each of the approaches in academia for a fair comparison. 


PAM~\cite{fowkes2016pam} is the only context-intensive approach, primarily designed for intra-project API pattern mining. In the paper, we also extend the approach to cross-project recommendation by selecting the best API from projects in the training set for each test case. The extended version of PAM is named as PAM-MAX, which indicates the theoretical maximum performance the context-insensitive approach can achieve.


\begin{table}
    \centering
    \caption{The evaluation metrics used in this empirical study.}
    \scalebox{0.95}{\begin{tabular}{ccp{2.5cm}}
    \toprule
         \textbf{Metric} & \textbf{Calculation} & \textbf{Description} \\
    \midrule
         \multirow{6}*{Success Rate@k} & \multirow{6}*{$\frac{\sum ^N _{i=1} \text{ContainCorrect}_k(q_i) }{N}$} & $\text{ContainCorrect}_k(q)$ returns 1 if the top k results returned for query $q$ contain the correct API, otherwise it returns 0. \\ 
         \midrule
         \multirow{4}*{MAP@k} & \multirow{4}*{\tabincell{c}{MAP@k = $\frac{\sum ^N _{i=1} AveP_k(q_i)}{N}$ \\ AveP@k = $\frac{\sum ^N _{i = 1} P_k(i) \times rel(i)}{m}$}} & $rel(i)$ returns 1 if the $i_{th}$ result is the correct API, otherwise it returns 0.\\
         \midrule
         \multirow{7}*{MRR} & \multirow{7}*{$\frac{\sum ^N _{i=1} 1/\text{firstpos}(q_i)}{N}$} & $\text{firstpos}(q)$ returns the position of the first correct API in the results, if it cannot find a correct API in results, it returns $+\infty$.   \\
         \midrule
         \multirow{13}*{NDCG@k} & \multirow{13}*{\tabincell{c}{NDCG@k = 
         $\frac{\sum ^N _{i=1}\frac{\text{DCG@k}(q_i)}{\text{IDCG}@k(q_i)}}{N}$ \\ \\ DCG@k = $\sum ^k _{t=1}\frac{rel_t(q_i)}{log_2(t+1)}$ }} & $rel_t(q)$ returns 2 if $t_{th}$ result exactly matches one correct API, and it returns 1 if $t_{th}$ result matches the API class but fails to match the API, otherwise it returns 0. IDCG@k is the best DCG@k by re-arranging the order of current results.\\
    \bottomrule
    \end{tabular}
}
    \label{tab:metrics}
\end{table}
\textbf{Evaluation metrics.} Since both query-based and code-based API recommendation baselines output a ranked list of candidate APIs, we adopt the commonly-used metrics in recommendation tasks for evaluation. Table~\ref{tab:metrics} shows the details of each metric.

The Mean Reciprocal Rank (MRR), Mean Average Precision (MAP), and Normalized Discounted Cumulative Gain (NDCG) metrics are widely adopted by previous API recommendation studies~\cite{huang2018api}. In this study, we also involve a new metric Success Rate. The Success Rate@k is defined to evaluate the ability of an approach in recommending correct APIs based on the top-k returned results regardless of the orders. To determine the relevance score in NDCG calculation, We use a relevance score of 1 if an approach hits the correct API class, and a relevance score of 2 if the correct API method is hit. Therefore, we can align the performance of class-level and method-level approaches.


\section{Empirical Results of Query Reformulation and Query Based API Recommendation}\label{sec:querybased}
In this section, we study the RQ 1-3 discussed in Sec.~\ref{sec:intro} and provide the potential findings concluded from the empirical experiments. Since currently no query-based API recommendation approach is specially designed for Python APIs, we focus on studying query-based API recommendation approaches for Java.

\begin{table*}[t]
    \centering
    \caption{The basic performance of query-based API recommendation baselines without applying any query reformulation techniques at different metrics (Top-1,3,5,10). Note that we define NDCG as a uniform metric to evaluate class level and method level together, so the NDCG scores listed in two levels have the same values. The \textcolor{red}{red} numbers indicate the best performance achieved in top-10 results.}
    \renewcommand\tabcolsep{5.5pt}
    \begin{tabular}{llccccccccccccc}
    \toprule
        \multirow{2}*{\textbf{Baseline}} & \multirow{2}*{\tabincell{c}{ \textbf{Level}}} & \multicolumn{4}{c}{\textbf{Success Rate@k}} & \multicolumn{4}{c}{\textbf{MAP@k}} & \multirow{2}*{\textbf{MRR}} & \multicolumn{4}{c}{\textbf{NDCG@k}} \\
        \cmidrule{3-10}
        \cmidrule{12-15}
        &  & Top-1 & Top-3 & Top-5 & Top-10 & Top-1 & Top-3 & Top-5 & Top-10 &  & Top-1 & Top-3 & Top-5 & Top-10 \\
        \midrule
        RACK & Class & 0.17 & 0.30 & 0.35 & 0.41 & 0.17 & 0.23 & 0.24 & 0.24 & 0.25 & 0.17 & 0.24 & 0.26 & 0.28 \\
        \specialrule{0em}{1pt}{1pt}
        KG-APISumm & Class & 0.19 & 0.33 & 0.40 & 0.50 & 0.19 & 0.25 & 0.26 & 0.27 & 0.28 & 0.19 & 0.24 & 0.27 & 0.31 \\
        \specialrule{0em}{1pt}{1pt}
        \multirow{2}*{Naive Baseline} & Class & 0.07 & 0.13 & 0.16 & 0.21 & 0.07 & 0.10 & 0.10 & 0.10 & 0.11 & 0.07 & 0.09 & 0.10 & 0.13 \\
         & \g Method & \g 0.02 & \g 0.03 & \g 0.04 & \g 0.05 & \g 0.01 & \g 0.02 & \g 0.03 & \g 0.03 & \g 0.03 & \g 0.07 & \g 0.09 & \g 0.10  & \g 0.13\\
        \specialrule{0em}{1pt}{1pt}
        \multirow{2}*{DeepAPI} & Class & 0.19 & 0.27 & 0.29 & 0.30 & 0.19 & 0.22 & 0.23 & 0.23 & 0.23 & 0.17 & 0.22 & 0.23 & 0.24 \\
         & \g Method & \g 0.05 & \g 0.09 & \g 0.10 & \g 0.11 & \g 0.05 & \g 0.07 & \g 0.07 & \g 0.07 & \g 0.07 & \g 0.17  & \g 0.22  & \g 0.23 & \g 0.24 \\
        \specialrule{0em}{1pt}{1pt}
        \multirow{2}*{Lucene} & Class & 0.15 & 0.21 & 0.24 & 0.29 & 0.15 & 0.17 & 0.18 & 0.17 & 0.19 & 0.12 & 0.15 & 0.16 & 0.20 \\
         & \g Method & \g 0.04 & \g 0.08 & \g 0.10 & \g 0.14 & \g 0.04 & \g 0.06 & \g 0.06 & \g 0.06 & \g 0.07 & \g 0.12  & \g 0.15  & \g 0.16  &  \g 0.20 \\
        \specialrule{0em}{1pt}{1pt}
        \multirow{2}*{BIKER} & Class & 0.33 & 0.51 & 0.59 & \textcolor{red}{\textbf{0.67}} & 0.33 & 0.41 & 0.41 & \textcolor{red}{\textbf{0.39}} & \textcolor{red}{\textbf{0.44}} & 0.27 & 0.32 & 0.35 & \textcolor{red}{\textbf{0.42}}\\
        & \g Method & \g 0.12 & \g 0.23 & \g 0.29 & \g \textcolor{red}{\textbf{0.37}} & \g 0.12 & \g 0.16 & \g 0.18 & \g \textcolor{red}{\textbf{0.18}} & \g \textcolor{red}{\textbf{0.19}} &\g 0.27  & \g 0.32  & \g 0.35  & \g \textcolor{red}{\textbf{0.42}} \\
    \bottomrule
    \end{tabular}
    \label{tab:querybaseres}
\end{table*}

\subsection{Effectiveness of Query-Based API Recommendation Approaches (RQ1-1)}\label{sec:rq1-1}
To answer RQ1, we evaluate the six query-based API recommendation baselines listed in Table~\ref{tab:querybaselines} by using the original queries in our benchmark \benchq. The evaluation results are illustrated in Table~\ref{tab:querybaseres}.


\textbf{Class-level v.s. Method-level.} Regarding the \textbf{class-level} recommendation, as shown in Table~\ref{tab:querybaseres}, we can find that BIKER achieves the highest Success Rate, e.g., 0.67 for Success Rate@10, indicating that BIKER is more effective in finding the correct API class in the top-10 returned results for 60\%$\sim$70\% of cases. Unsurprisingly, the naive baseline shows the worst performance for all the metrics. Even so, the naive baseline can successfully predict the correct API class for around 20\% of cases. However, with respect to the \textbf{method-level} recommendation, all the approaches show obvious declines. For example, the Success Rate@10 of BIKER is only 0.37, decreasing by 44.8\% compared to the class-level recommendation. The Success Rates@10 of DeepAPI and Lucene are only around 0.10, which is far from the requirement of practical development. On average, the approaches fail to give the exact methods for 57.8\% APIs that they give the correct classes in top-10 returned recommendations. Thus, recommending method-level APIs still remains a great challenge.


\finding{1}{Existing approaches fail to predict 57.8\% method-level APIs that could be successfully predicted at the class level. The performance achieved by the approaches is far from the requirement of practical usage. Accurately recommending the method-level APIs still remains a great challenge.}

\textbf{Retrieval-based methods v.s. Learning-based methods.} By comparing learning-based methods, such as DeepAPI and naive baseline, with the other retrieval-based methods, we can observe that learning-based methods achieve relatively lower performance regarding the Success Rate@10 metric. For example, on average, retrieval-based methods can accurately predict 46.8\% class-level and 25.5\%  method-level APIs among all the cases in the top-10 returned results, respectively, while learning-based methods can only successfully recommend 25.5\% class-level and 8\% method-level APIs. A possible reason may be the insufficient training data for the learning-based methods in this task domain. Since there are more than 30,000 APIs from the official documentation, learning-based methods require a large number of query-API pairs for training. However, even the largest Q\&A forum, Stack Overflow, contains only about 150,000 posts after our pre-processing, which is not enough for model training.


\finding{2}{Learning-based methods do not necessarily outperform retrieval-based methods in recommending more correct APIs. The insufficient query-API pairs for training limit the performance of learning-based methods.}

\textbf{Performance in API ranking.} From Table~\ref{tab:querybaseres}, we find that there exist obvious gaps between the scores of Success Rate@k and the metrics for evaluating API ranking, such as MAP@k and NDCG@k. For example, RACK achieves Success Rate@10 score at 0.41, but its MAP@10 score is only 0.24. This indicates that although the approaches are able to find the correct APIs, they cannot well rank them ahead in the returned results. The low MRR scores, e.g., 0.11 $\sim$ 0.44 for class-level API recommendation and 0.03 $\sim$ 0.19 for method-level API recommendation, and NDCG scores also show the poor ranking performance of the approaches. The results manifest that API ranking is still challenging for current approaches.



\finding{3}{Current approaches cannot well rank the correct APIs, considering the huge gap between the scores of Success Rate and the other ranking metrics.}

To sum up, accurately recommending method-level APIs and ranking candidate APIs still remain great challenges. Besides, the insufficient data for training hinder the performance of current learning-based approaches.


\subsection{Effectiveness of Query Reformulation Techniques (RQ2)}\label{sec:rq2}
Original queries can be short in length or contain vague terms. Query reformulation aims at changing original queries for facilitating downstream tasks. In this RQ, we explore the impact of query reformulation on the performance of query-based API recommendation.

We implement the four query reformulation techniques, as listed in Table~\ref{tab:querybaselines}, for the original queries. We name the queries reformulated by query expansion techniques and query modification techniques as expanded queries and modified queries, respectively. For each original query, we conduct the reformulation 10 times, producing 10 expanded or modified queries, with the statistics shown in Table~\ref{tab:benchq}. Note that NLPAUG~\cite{ma2019nlpaug} is a comprehensive data augmentation library for general NLP tasks. We choose the popular word-level insertion and substitution methods designed for manipulating single sentences based on five models, including BERTOverflow~\cite{bertovereflow}, Google News Word2vec~\cite{w2v}, Stack Overflow Word2vec~\cite{sow2v}, WordNet~\cite{wordnet}, and Random model, in the library to generate expanded and modified queries.


\begin{figure*}[htb]
    \centering
    \subfigure[Query Expansion]{
    \includegraphics[width = 0.9\textwidth]{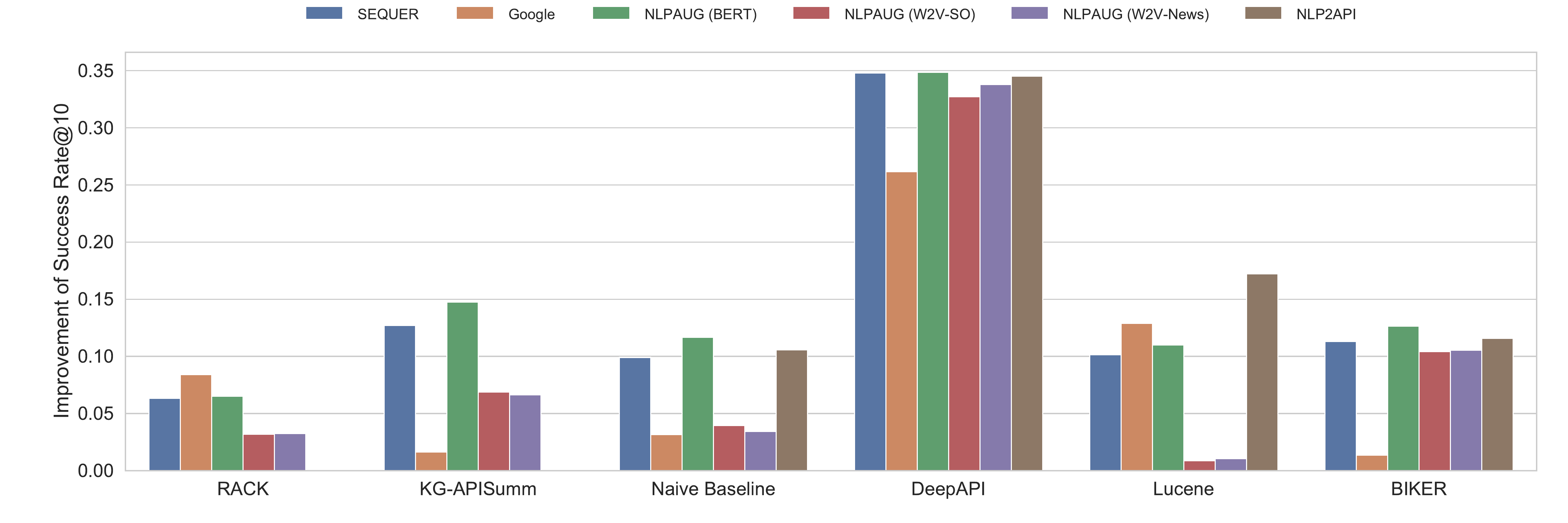}
    }
    \subfigure[Query Modification]{
    \includegraphics[width = 0.8\textwidth]{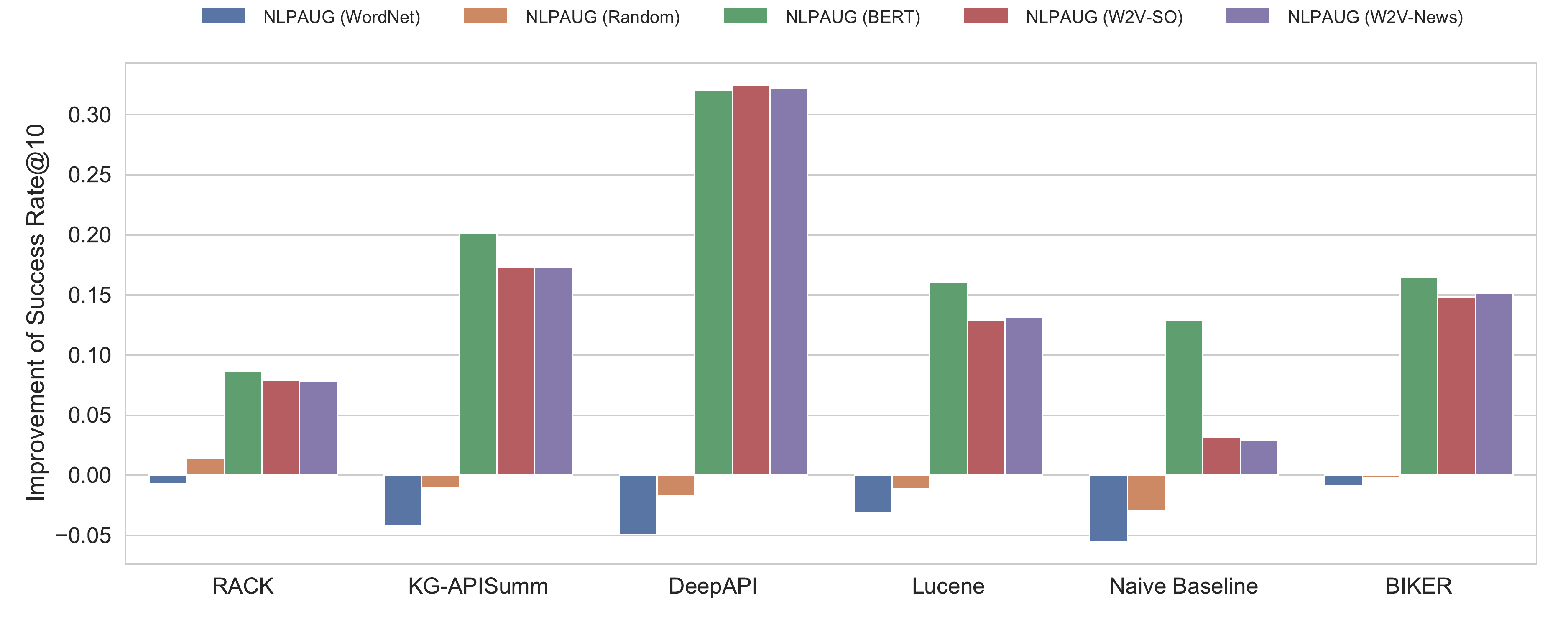}
    }
    \caption{The maximum improvement of Success Rate@10 by all query reformulation techniques on \textbf{class-level} query-based API recommendation baselines. We do not evaluate the performance of RACK and KG-APISumm in NLP2API reformulated queries as they are only class-level recommendation approaches while NLP2API directly give the predicted API classes. Note that we include Google Prediction Service and SEQUER as expansion techniques here because they expand the queries in most cases.}
    \label{fig:reformclassres}
\end{figure*}

The queries output by the query reformulation techniques are not ranked in order, and may impact the downstream API recommendation performance variously. To explore the maximum potential effect brought by query reformulation techniques, we evaluate the API recommendation approaches on each reformulated query and choose the best result for analysis. For example, SEQUER~\cite{cao2021automated} generates 10 queries for the original query ``\textit{Get min value between two double type values}". We choose the one achieving the best performance for analysis, i.e., using the reformulated query ``\textit{Get min value between two double type values in java}" for the recommendation. We study the impact of query reformulation on API recommendation from the following two aspects:

1) whether query reformulation techniques can help predict more correct APIs;

2) whether query reformulation can improve the API ranking performance.




\subsubsection{Influence on predicting more correct APIs}
\begin{figure*}[t]
    \centering
    \subfigure[Query Expansion]{
    \includegraphics[width = 0.85\textwidth]{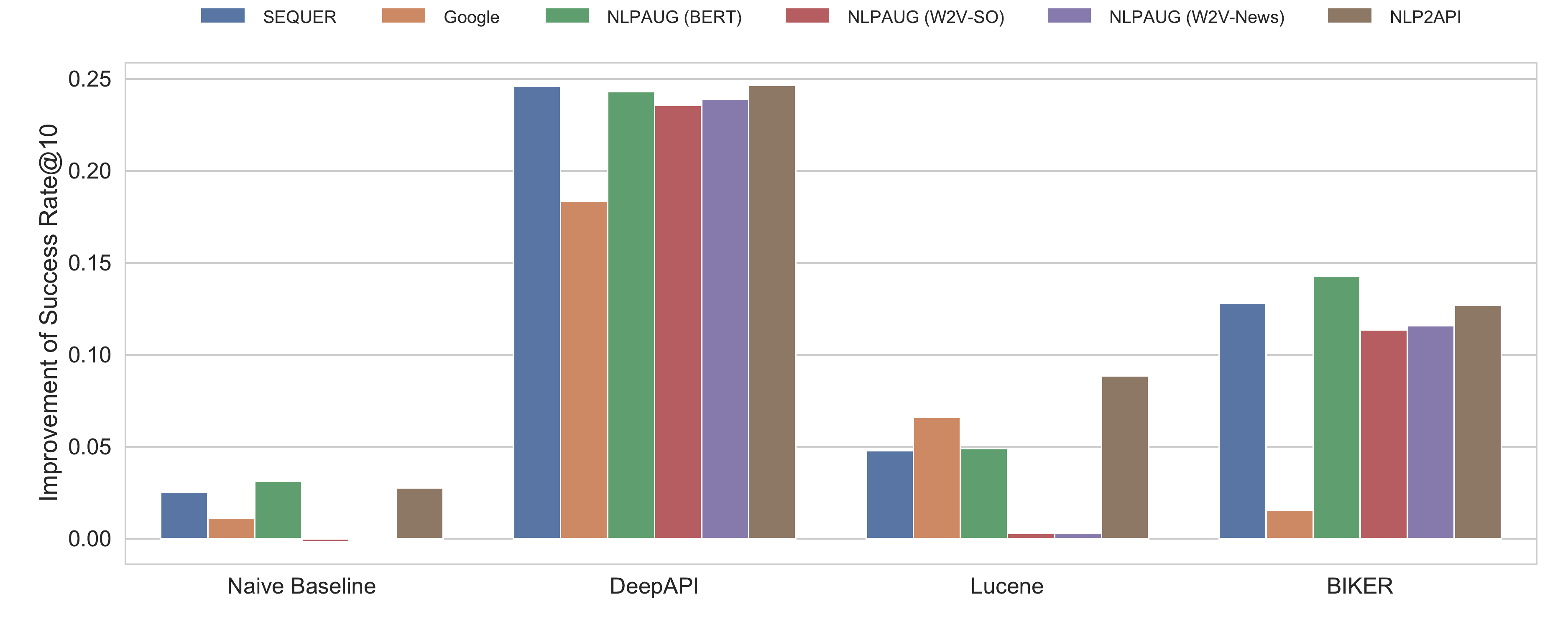}
    }
    \subfigure[Query Modification]{
    \includegraphics[width = 0.7\textwidth]{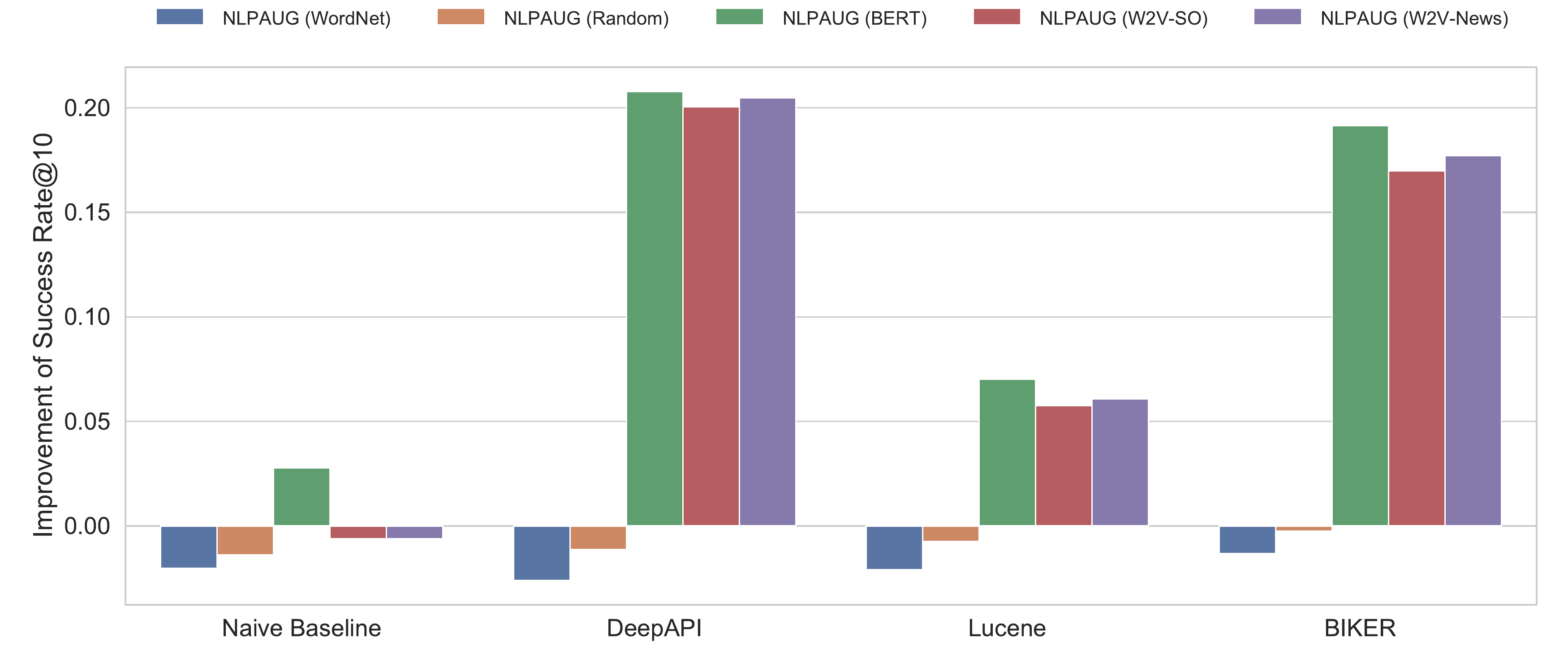}
    }
    \caption{The maximum improvement of Success Rate@10 by all query reformulation techniques on \textbf{method-level} query-based API recommendation baselines.}
    \label{fig:reformapires}
\end{figure*}

\textbf{With query reformulation v.s. Without query reformulation.} The Success Rate metric reflects the proportion of the APIs an approach can correctly predict. The results of implementing the reformulation techniques on API recommendation approaches are illustrated in Figure~\ref{fig:reformclassres} (class-level) and Figure~\ref{fig:reformapires} (method-level). From the figures, we observe that query reformulation can increase the performance of API recommendation in most cases. Only for a few cases, the performance drops, which can be attributed to the inefficiency of some query reformulation techniques. For example, NLPAUG (WordNet) and NLPAUG (Random) tend to poorly modify the original queries for recommendation, as shown in Figure~\ref{fig:reformclassres} (b) and Figure~\ref{fig:reformapires} (b). Overall, on average the process improves the class-level and method-level recommendation by 0.11 and 0.08, which is a boost of 27.7\% and 49.2\% compared with the basic performance on original queries. 


\finding{4}{Query reformulation techniques, including query expansion and query modification, are quite effective in helping query-based API recommendation approaches give the correct API by adding an average boost of 27.7\% and 49.2\% on class-level and method-level recommendations. }

\textbf{Query expansion v.s. Query modification.} By comparing the class-level and method-level recommendation results of query expansion and query modification in Figure~\ref{fig:reformclassres} and Figure~\ref{fig:reformapires}, respectively, We observe that all the query expansion techniques improve the API recommendation performance, but not all the query modification techniques benefit the recommendation. For example, NLPAUG (WordNet) and NLPAUG (Random) generally decrease the performance of current approaches both in class-level and method-level recommendations. This indicates that query expansion techniques bring more stable improvement than query modification techniques. Furthermore, on average, query expansion techniques improve the performance by 0.13 and 0.10 on class-level and method-level recommendation, which is much higher than the improvement of 0.09 and 0.06 achieved by query modification techniques. This also suggests that query expansion techniques are more effective than query modification techniques.


\finding{5}{Query expansion is more stable and effective to help current query-based API recommendation approaches give correct APIs than query modification.}

\textbf{Comparing different query expansion techniques.} As shown in Figure~\ref{fig:reformclassres} (a) and Figure~\ref{fig:reformapires} (b), NLP2API and NLPAUG (BERT) present the largest improvement on the performance of query-based API approaches at both class level and method level. For analyzing the improvement, we use two examples to illustrate the query expansion results of NLP2API and NLPAUG (BERT), respectively. In both examples, the most effective approach BIKER fails to predict the API based on the original queries but succeeds given the reformulated queries.

\example{1}{Query Expansion}{NLP2API}{Returns a new Document instance}{\colorbox{mygray}{DocumentBuilderFactory} Returns a new Document instance}

In the first example, NLP2API expands the query by adding a predicted API class \textit{DocumentBuilderFactory} that is related to the original query. With such an explicit hint, the recommendation approach can narrow down the search scope and pinpoint the requested API method.

\example{2}{Query Expansion}{NLPAUG (BERT)}{Java reverse string}{java reverse \colorbox{mygray}{character} string}

In the second example, the query is looking for the API \textit{java.lang.StringBuilder.reverse()}, whose description in official documentation is ``\textit{Causes this character sequence to be replaced by the reverse of the sequence}''. NLPAUG (BERT) adds a relevant word \textit{character} to enrich the semantics of the original query. 

\example{3}{Query Expansion}{NLPAUG (W2V-SO)}{Convert from Radians to Degrees in Java}{Convert from \colorbox{mygray}{AV} Radians to Degrees in \colorbox{mygray}{Long} Java}

Comparing NLPAUG (W2V) with NLPAUG (BERT) and NLP2API in Figure~\ref{fig:reformclassres} and Figure~\ref{fig:reformapires}, we find that the NLPAUG (W2V) is much less effective. To obtain a possible reason for such a difference, we give the third example below. As shown in the third example,  we find that NLPAUG (W2V) adds two irrelevant words into the original query, which negatively impacts the prediction results of BIKER. This also indicates that contextual embeddings such as BERT are more effective than traditional word embeddings.

\finding{6}{Add predicted API class names or relevant words into original queries are effective query expansion methods for improving the performance of API recommendation.}

\textbf{Comparing different query modification techniques.}  Among all query modification techniques, NLPAUG (BERT) presents the biggest improvement on all the baselines at both class level and method level. Example 4 illustrates how NLPAUG (BERT) modifies words in the original query. In the example, the original query
asks about ways to calculate the time difference between two dates and the correct API is \textit{java.time.Period.between()}. The description of the API in its official documentation is ``\textit{obtains a period consisting of the number of years, months, and days between two dates}''. However, the word ``\textit{difference}'' used in the original query does not clearly describe the functional request. NLPAUG (BERT) modifies the word into ``\textit{months}'' which exactly appears in the official description. Based on the modifications, the correct API is recommended. 


\example{4}{Query Modification}{NLPAUG (BERT)}{How do I calculate difference between two dates}{how do \colorbox{mygray}{they \sout{I}} calculate \colorbox{mygray}{months \sout{difference}} between two dates}

From the second and fourth examples above, we find that BERT-based models show great performance on both query expansion and query modification to help improve the performance of current query-based API recommendation approaches. This indicates that even though the current data source limits the performance of them to directly predict the correct APIs, they can be used to improve the query quality as query reformulation techniques.


\finding{7}{BERT-based data augmentation shows superior performance in query modification compared with other query modification techniques.}

\begin{figure*}[t]
    \centering
    \subfigure[Query Expansion]{
    \includegraphics[width = 1.0\textwidth]{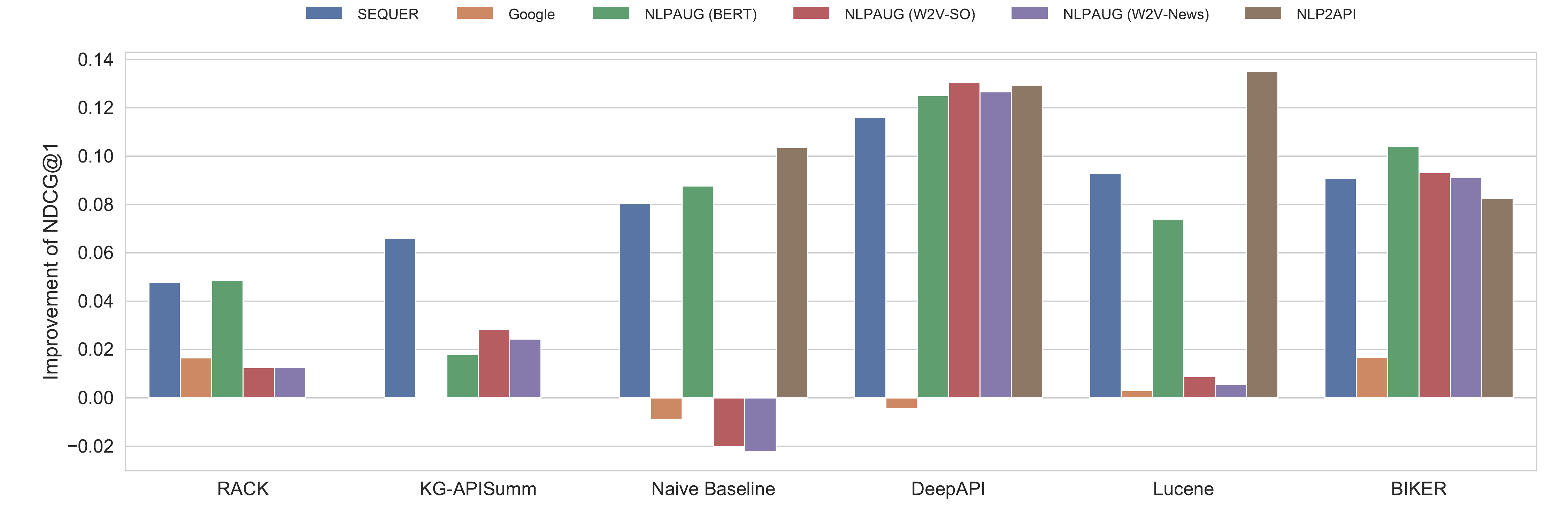}
    }
    \subfigure[Query Modification]{
    \includegraphics[width = 0.8\textwidth]{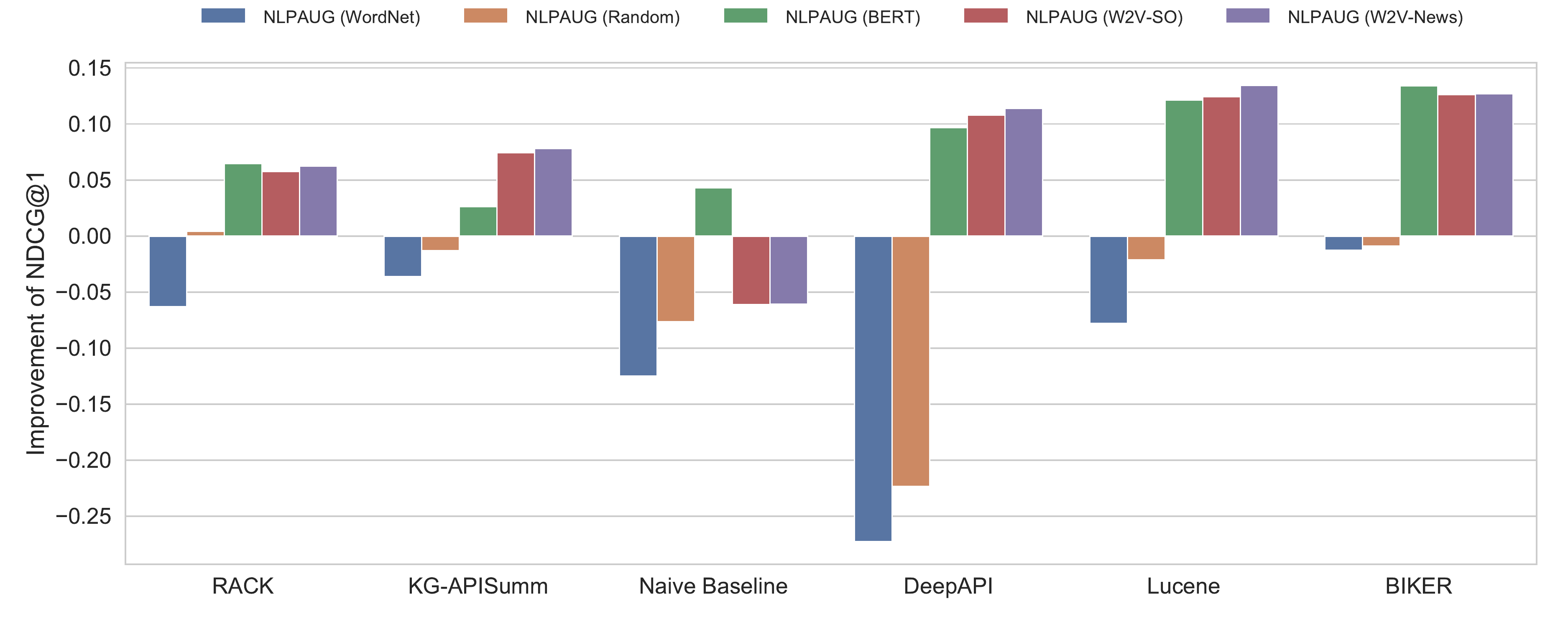}
    }
    \caption{The maximum improvement of NDCG@1 by all query reformulation techniques on query-based API recommendation baselines under original successful cases.}
    \label{fig:reformresndcg}
\end{figure*}

\subsubsection{Influence on the performance of API ranking}
In this section, we analyze the impact of query reformulation techniques on the performance of API ranking. Since the ideal case is that the correct APIs rank first in the returned results, we use the metric NDCG@1 which considers both class-level and method-level recommendation performance. We compute the changes of NDCG@1 scores for the query-based API recommendation approaches before and after query reformulation. Besides, to focus our analysis on the performance of API ranking instead of the overall recommendation accuracy, the computation is performed only on the cases that are correctly predicted with and without query reformulation.


The results are illustrated in Figure~\ref{fig:reformresndcg}. As can be seen in Figure~\ref{fig:reformresndcg} (a), most query expansion techniques also improve the ranking results of the query-based recommendation approaches. Among all the query expansion techniques, SEQUER, NLPAUG (BERT), RACK and NLP2API can relatively better improve the ordering performance. The biggest improvement 0.14 (32\% boost) is achieved by NLP2API on the Lucene approach. We also find that on average query expansion also improves MRR by 0.09 (36\% boost) and 0.08 (89\% boost) on class-level and method-level recommendation, which indicates that the correct APIs are ranked much higher based on reformulated queries.

According to Figure~\ref{fig:reformresndcg} (b), compared with query expansion techniques, query modification techniques are much less effective in improving the API ranking performance. For example, the average improvement of NDCG@1 brought by query modification is 0.01 (4\% boost), which is 0.06 (14\% boost) for query expansion techniques. Comparing different data augmentation methods, we also find that WordNet and random methods tend to negatively impact the ranking results, leading to 24\% and 14\% drop in terms of NDCG@1, respectively. The results indicate that inappropriate query modification will reduce the ranking performance of the query-based recommendation approaches.

\finding{8}{Expanding queries or modifying queries with appropriate data augmentation methods can improve the ranking performance of the query-based API recommendation techniques.}

To sum up, query reformulation, especially query expansion, can not only help current approaches recommend more correct APIs, but also improve the ranking performance. However, the reformulation step is generally ignored by current studies. Future work is suggested to involve such a step for more accurate API recommendation.

\begin{figure}[t]
    \centering
    \subfigure[Class Level]{
    \includegraphics[width = 0.5\textwidth]{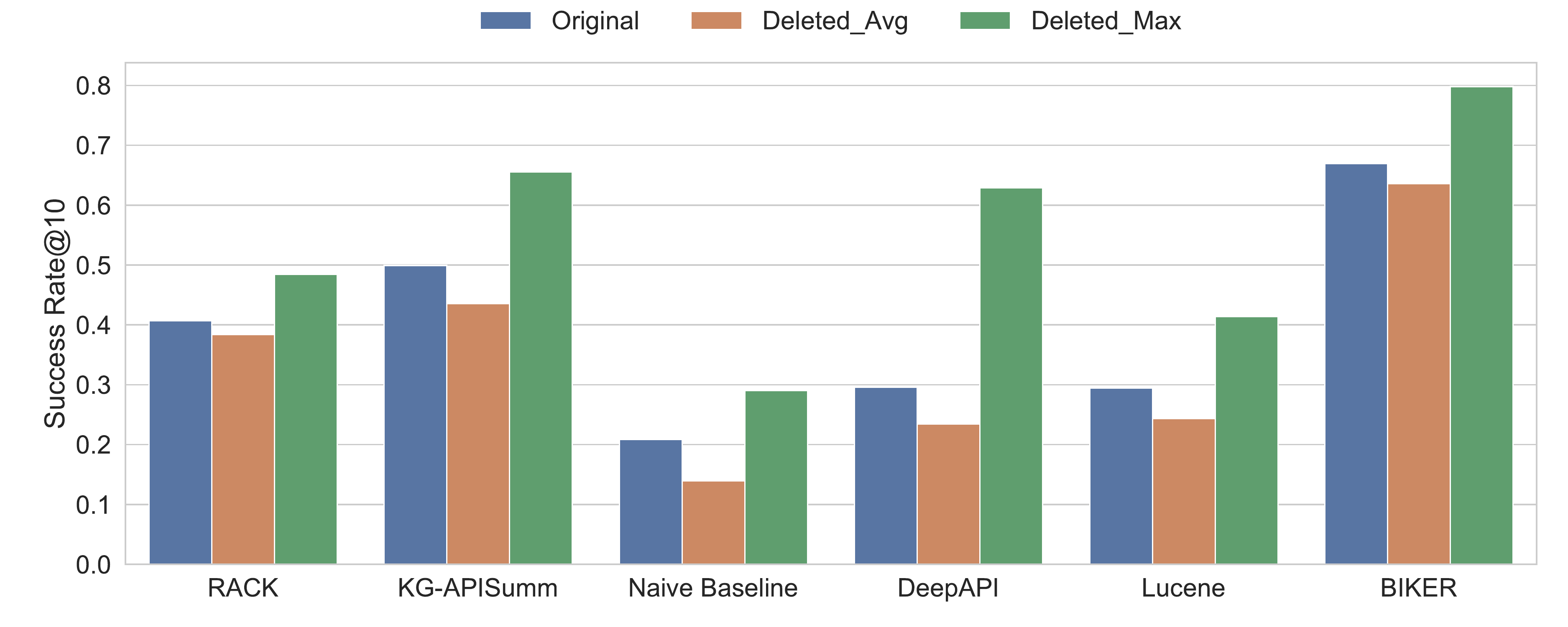}
    }
    \subfigure[Method Level]{
    \includegraphics[width = 0.4\textwidth]{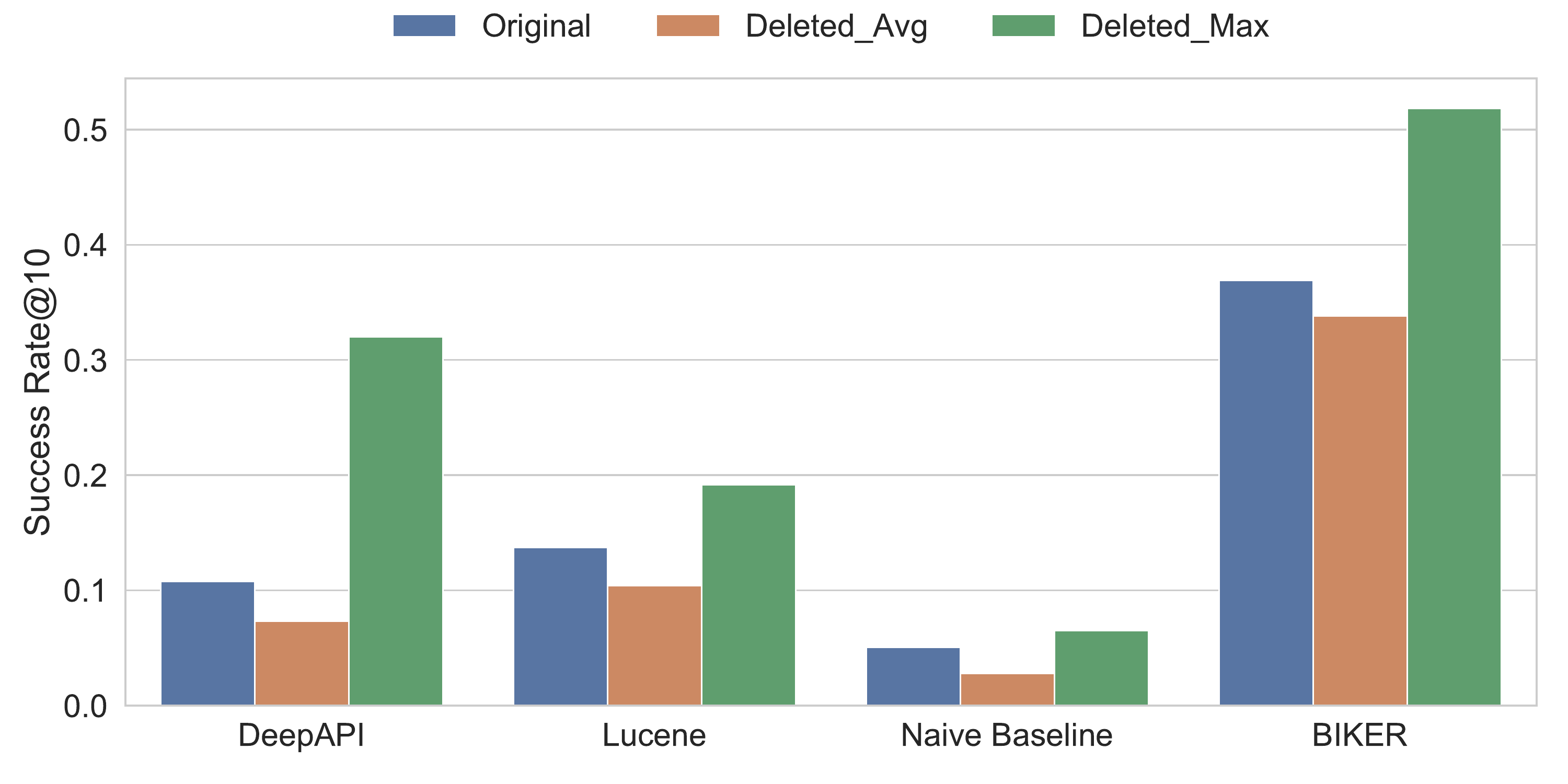}
    }
    \caption{The maximum and average Success Rate@10 on all baselines when randomly deleting some words in original queries.}
    \label{fig:deleteres}
\end{figure}

\subsubsection{A special Query Modification Method: Word Deletion}\label{sec:del}

In previous subsections, we compare and evaluate different query expansion and modification techniques. They aim at enriching the original queries by adding, replacing or modifying some words without deleting words. In this section, we focus on studying the impact of word deletion, a special query modification method, on the performance of query-based API recommendation approaches. Different from the previous query reformulation techniques which rely on external data sources, the word deletion method we studied does not leverage any extra knowledge. Our goal is to explore whether original queries contain meaningless or noisy words. Specifically, we randomly delete some words from the original query every time and produce ten different modified queries for one original query.

The maximum and average Success Rate@10 scores based on modified queries are illustrated in Figure~\ref{fig:deleteres}. As can be seen, the average performance of the query-based API recommendation approaches, denoted as the orange bar, decreases by 0.05 (13\% drop) at class level and 0.03 (18\% drop) at method level. The results are not surprising, and indicate that most words in the original queries are helpful for the recommendation. However, the maximum scores, denoted as the green bar, all show that word deletion improves the recommendation performance with an average boost of 38\% and 64\% for class level and method level, respectively. The improvement demonstrates that the original queries contain noisy words that can bias the recommendation results, although most of the words are useful for recommendation.

To understand what kinds of words are noisy for the accurate recommendation, we manually check 545 out of 6,563 queries for which the recommendation approaches perform better after word deletion. We summarize the cases as below:

1) 349 (64\%) queries contain unnecessary or meaningless words. 

\example{5}{Word Deletion}{Random Deletion}{Standard way to iterate over a StringBuilder in java}{\colorbox{mygray}{\sout{Standard way to}} iterate over a StringBuilder \colorbox{mygray}{\sout{in java}}}

In Example 5, the phrases ``\textit{Standard way to}'' and ``\textit{in java}'' are not beneficial for pinpointing the correct API. Stop word removal also has a limited effect on eliminating these words.
    
2) 156 (29\%) queries contain too detailed words for explanation.
\example{6}{Word Deletion}{Random Deletion}{converts a color into a string like 255,0,0}{converts a color into a string \colorbox{mygray}{\sout{like 255,0,0}}}

In Example 6, the phrase ``\textit{like 255,0,0}'' is used to explain the ``\textit{string}''. However, such phrases never appear in the official documentation and the specific number adversely impacts the recommendation results.

3) 34 (6\%) queries contain extreme long descriptions. 
\example{7}{Word Deletion}{Random Deletion}{how to add progress bar to zip utility while zipping or extracting in java}{how to add progress bar to zip utility \colorbox{mygray}{\sout{while zipping or extracting in java}}}

In Example 7, the words after while actually describe nothing about the task. The long descriptions can decrease the weight of useful words in the queries thus confusing API recommendation approaches.

\finding{9}{Original queries raised by users usually contain noisy words which can bias the recommendation results, and query reformulation techniques should consider involving noisy word deletion for a more accurate recommendation.}

\subsection{Data Sources (RQ3)}\label{sec:datasource}
In RQ1-1, we highlight that insufficient data greatly limits the performance of current learning-based methods. In this section, we conduct a deep analysis on the influence of different data sources on the recommendation results. From Table~\ref{tab:querybaselines}, we can observe that current approaches generally leverage three different data sources: official documentation, Q\&A forums, and tutorial websites. For analysis, we choose two methods, Lucene and naive baseline, which are flexible to incorporate different data sources. Specifically, we evaluate the methods on the part of queries from the tutorial websites collected in \benchq, and the method training is conducted based on the following knowledge base:

1) only official documentation, 

2) only Stack Overflow posts, and 

3) both official documentation and Stack Overflow posts.


\begin{figure}[t]
    \centering
    \includegraphics[width = 0.8\linewidth]{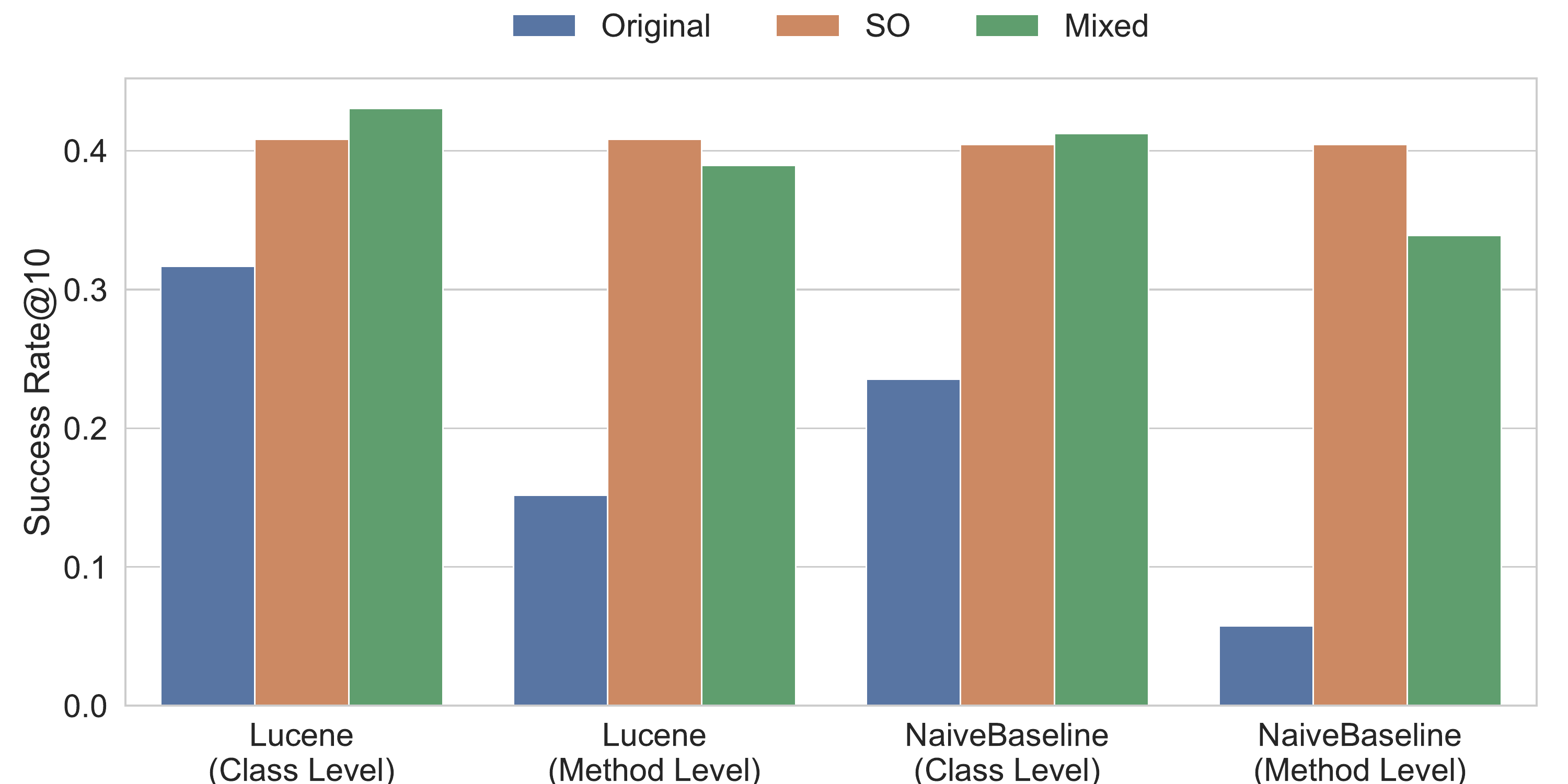}
    \caption{The Success Rate@10 of Lucene and Naive Baseline under three data source settings.}
    \label{fig:datasource}
\end{figure}

The experiment results are shown in Figure~\ref{fig:datasource}. As can be seen, training on Stack Overflow posts achieves much better performance than on official documentation at both class and method levels. For example, Lucene achieves a 29\% boost in class-level and an 169\% boost in method-level recommendation when searching based on Stack Overflow than on official documentation; and the naive baseline even achieves a 71\% boost in class-level and a 602\% boost in method-level recommendation. The advantage of leveraging Stack Overflow posts may be attributed that the discussion on Stack Overflow is more natural and similar to user queries, compared with the descriptions in the official documentation. Besides, the extended usage of some APIs is rarely mentioned in official documentation but is widely discussed in Stack Overflow. An example is used to illustrate the influence of different data sources.

\begin{table}[ht]
    \centering
    \begin{tabular}{lp{5.7cm}}
    \toprule
    \rowcolor{mygray}
    \multicolumn{2}{l}{\textbf{Example 8: Data Source}} \\
    \textsc{Baseline} & Lucene \\
    \textsc{Original Query} & Compute the md5 hash of a File  \\
    \textsc{Correct API} & java.security.MessageDigest.digest(), java.security.MessageDigest.getInstance()  \\
    \textsc{API Description} & Completes the hash computation by performing final operations such as padding \\
    \textsc{Similar SO Post} & How can I generate an MD5 hash in Java? \\
    \bottomrule
    \end{tabular}
\end{table}

In Example 8, the query asks about the API for generating an MD5 hash of a file. However, there is no standard API specially designed to generate the MD5 hash, so Lucene focuses on two words ``\textit{hash}" and ``\textit{file}" for recommendation. But the official description of ground truth API \textit{java.security.MessageDigest.digest()} does not contain the word ``\textit{file}" since it is a general API that not only handles files. Under this circumstance, Lucene recommends a more relevant but wrong API \textit{java.nio.file.attribute.FileTime.hashCode()}. When involving Stack overflow posts, as there already exists discussion on how to generate the MD5 hash, Lucene can easily pinpoint and recommend the API in the posts.

The advantage of leveraging Stack Overflow for recommendation is also demonstrated by the BIKER approach~\cite{huang2018api}, which is the most effective approach in Section~\ref{sec:rq1-1}. Our finding is consistent with the claim in the work~\cite{huang2018api} that Stack Overflow posts can mitigate the semantics gap between user queries and official descriptions.


\finding{10}{Apart from official documentation, using other data sources such as Stack Overflow can significantly improve the performance of query-based API recommendation approaches.}

\begin{table*}
    \centering
    \caption{The performance of code-based API recommendation baselines at different metrics (Top-1,3,5,10). All baselines are trained and tested on the full dataset from ``General'' domain of \benchc except for PyART. Since PyART takes months to train and test on our full dataset, we randomly sampled 20\% of original training and testing testset to evaluate it.  The ``PL'' column indicates the programming language the baselines target. The \textcolor{red}{red} number indicates the best performance.}
    \renewcommand\tabcolsep{5.5pt}
    \begin{tabular}{cccccccccccccccc}
    \toprule
        \multirow{2}*{\textbf{PL}} & \multirow{2}*{\textbf{Baseline}} & \multicolumn{4}{c}{\textbf{Success Rate@k}} & \multicolumn{4}{c}{\textbf{MAP@k}} & \multirow{2}*{\textbf{MRR}} & \multicolumn{4}{c}{\textbf{NDCG@k}} \\
        \cmidrule{3-10}
        \cmidrule{12-15}
        & & Top-1 & Top-3 & Top-5 & Top-10 & Top-1 & Top-3 & Top-5 & Top-10 &  & Top-1  & Top-3 & Top-5 & Top-10 \\
        \midrule
        \specialrule{0em}{1pt}{1pt}
        \multirow{3}*{Python} & TravTrans & 0.45 & 0.57 & 0.59 & \textcolor{red}{\textbf{0.62}} & 0.45 &  0.50 & 0.51 & \textcolor{red}{\textbf{0.51}} & \textcolor{red}{\textbf{0.51}} & 0.45  & 0.52 & 0.53 & \textcolor{red}{\textbf{0.54}} \\
        \specialrule{0em}{1pt}{1pt}
        & Deep3 & 0.21 & 0.34 & 0.37 & 0.43 & 0.20 & 0.27 & 0.28 & 0.28 & 0.28 & 0.21 & 0.29 & 0.30 & 0.32 \\
        \specialrule{0em}{1pt}{1pt}
        & PyART & 0.29 & 0.38 & 0.46 & 0.60 & 0.29 & 0.33 & 0.35 & 0.37 & 0.37 & 0.29 & 0.34 & 0.37 & 0.41 \\
        \midrule
        \multirow{3}*{Java} & FOCUS &  0.01 & 0.03 & 0.04 & 0.06 & 0.01 & 0.02 & 0.02 & 0.03 & 0.03 & 0.01 & 0.02 & 0.03 & 0.04 \\
        \specialrule{0em}{1pt}{1pt}
        & PAM & 0.01 & 0.02 & 0.03 & 0.05 & 0.01 & 0.02 & 0.02 & 0.02 & 0.02 & 0.01 & 0.02 & 0.02 & 0.03 \\
        \specialrule{0em}{1pt}{1pt}
        & PAM-MAX & 0.22 & 0.32 & 0.36 & \textcolor{red}{\textbf{0.45}} & 0.22 & 0.26 & 0.27 & \textcolor{red}{\textbf{0.28}} & \textcolor{red}{\textbf{0.28}} & 0.22 & 0.27 & 0.29 & \textcolor{red}{\textbf{0.32}} \\
    \bottomrule
    \end{tabular}
    \label{tab:codebaseres}
\end{table*}

\begin{table*}[t]
    \centering
    \caption{The performance of code-based API recommendation baselines along with 4 widely used IDEs tested on 500 cases sampled from the testset of all domains in \benchc. The ``PL'' column indicates the programming language the baselines target. The \textcolor{red}{red} number indicates the best performance. The rows with \colorbox{mygray}{gray} background indicates the performance of IDEs.}
    \renewcommand\tabcolsep{5.5pt}
    \begin{tabular}{cccccccccccccccc}
     \toprule
        \multirow{2}*{\textbf{PL}}& \multirow{2}*{\textbf{Baseline}} & \multicolumn{4}{c}{\textbf{Success Rate@k}} & \multicolumn{4}{c}{\textbf{MAP@k}} & \multirow{2}*{\textbf{MRR}} & \multicolumn{4}{c}{\textbf{NDCG@k}} \\
        \cmidrule{3-10}
        \cmidrule{12-15}
        & & Top-1 & Top-3 & Top-5 & Top-10 & Top-1 & Top-3 & Top-5 & Top-10 &  & Top-1  & Top-3 & Top-5 & Top-10 \\
        \midrule
        \multirow{4}*{Python} & TravTrans & 0.38 & 0.46 & 0.48 & \textcolor{red}{\textbf{0.50}} & 0.38 & 0.42 & 0.42 & \textcolor{red}{\textbf{0.43}} & \textcolor{red}{\textbf{0.43}} &  0.38 & 0.43 & 0.44 &\textcolor{red}{\textbf{0.44}}  \\
        \specialrule{0em}{1pt}{1pt}
        & Deep3 & 0.19 & 0.26 & 0.31 & 0.38 & 0.19 & 0.22 & 0.23 & 0.24 & 0.24 & 0.19 & 0.23  & 0.25 & 0.28  \\
        \specialrule{0em}{1pt}{1pt}
        & \g PyCharm & \g 0.31 & \g 0.42 & \g 0.47 & \g 0.49 & \g 0.31 & \g 0.36 & \g 0.37 & \g 0.37 & \g 0.37 & \g 0.31 & \g 0.38 & \g 0.40 & \g 0.40 \\
        \specialrule{0em}{1pt}{1pt}
        & \g VSCode & \g 0.05 & \g 0.15 & \g 0.21 & \g 0.35 & \g 0.05 & \g 0.09 & \g 0.11 & \g 0.13 & \g 0.13 & \g 0.05 & \g 0.11 & \g 0.14 & \g 0.18 \\
        \midrule
        \multirow{5}*{Java} & FOCUS & 0.02 & 0.04 & 0.05 & 0.07 & 0.02 & 0.03 & 0.03 & 0.04 & 0.04 & 0.02 & 0.03 & 0.04 & 0.04 \\
        \specialrule{0em}{1pt}{1pt}
        & PAM & 0.01 & 0.02 & 0.05 & 0.07 & 0.01 & 0.02 & 0.02 & 0.03 & 0.03 & 0.01  & 0.02 & 0.03  & 0.04  \\
        \specialrule{0em}{1pt}{1pt}
        & PAM-MAX & 0.27  & 0.38 & 0.43 & 0.56 & 0.27 & 0.31 & 0.33 & 0.34 & 0.34 & 0.27 & 0.33  & 0.35 & 0.39  \\
        \specialrule{0em}{1pt}{1pt}
        & \g Eclipse & \g 0.28 & \g 0.42 & \g 0.49 & \g 0.60 & \g 0.28 & \g 0.34 & \g 0.35 & \g 0.37 & \g 0.37 & \g 0.28 & \g 0.36 & \g 0.39 & \g 0.42 \\
        \specialrule{0em}{1pt}{1pt}
        & \g IntelliJ IDEA & \g 0.42  & \g 0.58 & \g 0.65 & \g \textcolor{red}{\textbf{0.67}} & \g 0.42 & \g 0.49 & \g 0.51 & \g \textcolor{red}{\textbf{0.51}} & \g \textcolor{red}{\textbf{0.51}} & \g 0.42 & \g 0.51 & \g 0.54 & \g \textcolor{red}{\textbf{0.55}} \\
        \specialrule{0em}{1pt}{1pt}
    \bottomrule
    \end{tabular}
    \label{tab:codesampleres}
\end{table*}

\section{Empirical Results of Code-Based API Recommendation}\label{sec:codebased}

In this section, we study the RQ1 and RQ 4 $\sim$ 6 discussed in Sec~\ref{sec:intro}. To study RQ1, RQ4 and RQ5, we evaluate the performance of all the code-based API recommendation approaches on the ``General'' domain of our benchmark \benchc, as shown in Table~\ref{tab:benchc}, since the ``General'' domain includes code with different topics and can reflect the overall performance of baselines. For studying the ability of cross-domain adaptation in RQ6, we evaluate the performance of the approaches on all the five domains of our \benchc.

\subsection{Effectiveness of Existing Approaches (RQ1-2)}\label{sec:rq1-2}
According to Table~\ref{tab:codebaselines}, three approaches for Python and three approaches for Java are evaluated on the ``General'' domain of \benchc. The results are depicted in Table~\ref{tab:codebaseres}. We can observe that the learning-based method TravTrans obtains the best performance on the Python dataset, achieving 0.62 and 0.54 for Success Rate@10 and NDCG@10, respectively. The results mean that TravTrans can successfully recommend 62\% of APIs in our benchmark and well predict the API rankings. However, the traditional statistical method Deep3 only achieves 0.43 and 0.32 for Success Rate@10 and NDCG@10, respectively, while the pattern-based method FOCUS and PAM achieve less than 0.10 for both Success Rate@10 and NDCG@10. This suggests that learning-based methods obtain superior performance in code-based API recommendation, which is quite different from query-based API recommendation. The possible reason is that lots of well-organized public code repositories provide sufficient data for training code-based API recommendation models. 

We also find that FOCUS and PAM show low recommendation accuracy, with all the metric values lower than 0.1. The low performance is attributed to the context representation of the approaches. PAM is a context-insensitive approach, which only mines the top-N APIs that are most likely to be used in the training set and directly recommends them for each file in the test set; while FOCUS takes one step further by extracting the APIs in the test set and building a matrix to match the APIs in the training set. Such coarse-grained context representation or context-insensitive representation do not well capture the relations between APIs. PAM-MAX shows the theoretical best performance context-insensitive methods can achieve. However, the performance of PAM-MAX is still lower than TravTrans and PyART which considers fine-grained code features such as code tokens and data flows. The results indicate the effectiveness of fine-grained approaches for code-based API recommendation.

Besides the recent code-based API recommendation approaches, we also compare the widely-used IDEs. Since it is hard to automatically evaluate IDEs' recommendation performance, we sampled 500 APIs from the original large test set of \benchc based on the distribution shown in Table~\ref{tab:benchc}. We then conduct a manual evaluation by imitating the behaviors of developers on the 500 sampled APIs. We show the results on the sampled test set in Table~\ref{tab:codebaseres}. As can be seen, for Python, Pycharm achieves the Success rate@10 at 0.49 and NDCG@10 at 0.40, which is truly competitive to the performance of TravTrans, with Success Rate@10 and NDCG@10 at 0.50 and 0.44, respectively. For Java, IDEs also show competitive performance compared with the baseline approaches. The results demonstrate that the widely-used IDEs are generally effective in API recommendation and far from relying on alphabet orders for recommendation.

\finding{11}{DL models such as TravTrans show superior performance on code-based API recommendation by achieving a Success Rate@10 of 0.62, while widely-used IDEs also obtain satisfying performance by achieving a Success Rate@10 of 0.5 $\sim$ 0.6.}

\subsection{Capability to Recommend Different Kinds of APIs (RQ4)}\label{sec:rq4}

\begin{figure}[t]
    \centering
    \includegraphics[width = 0.5\textwidth]{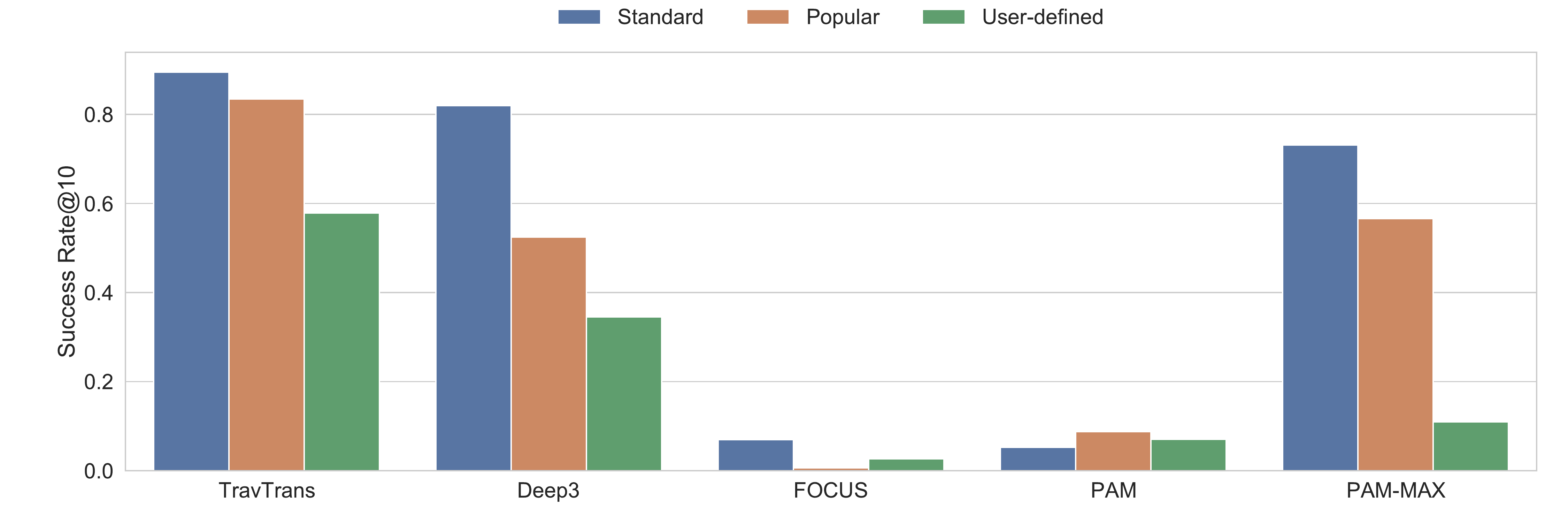}
    \caption{The Success Rate@10 of baselines on three categories of APIs at the ``{General}'' domain of \benchc.}
    \label{fig:stduserpop}
\end{figure}

Exploring which kinds of APIs tend to be wrongly predicted is essential for understanding the bottleneck of current approaches and providing clues for further improvement.
In Section~\ref{sec:methodology}, we have classified all APIs into standard APIs, popular third-party APIs and user-defined APIs. In this RQ, we study the performance of current baselines for different kinds of APIs. Specifically, we evaluate TravTrans, Deep3, FOCUS, PAM and PAM-MAX on the full test set of the ``General'' domain, with results shown in Figure~\ref{fig:stduserpop}. Note that we do not involve IDEs in this RQ as they are evaluated on the sampled test set.

As can be seen in Figure~\ref{fig:stduserpop}, all the approaches achieve a very high Success Rate@10 on standard APIs. For example, TravTrans even successfully recommends more than 90\% of standard APIs in the test set.
The approaches also present relatively good performance for the popular third-party libraries, e.g., TravTrans achieves
a Success Rate@10 of more than 0.8.
As standard APIs and popular APIs from third-party libraries are widely used in real-world projects, data-driven methods can achieve superior performance. However, the approaches are hard to correctly recommend the user-defined APIs and fail to predict 35.3\% $\sim$ 91.3\% more of user-defined APIs comparing to the prediction of standard APIs.

\finding{12}{Although current approaches achieve good performance on recommending standard and popular third-party libraries, they face the challenges of correctly predicting the user-defined APIs.}

\subsection{Capability to Handle Different Contexts (RQ5)}\label{sec:rq5}

As context representation is an important part of the current code-based API recommendation shown in Figure~\ref{fig:codemodel}, it is worthwhile to study the impact of different contexts on the performance of current approaches. In this RQ, we explore the impact of the following two different types of context.

\begin{itemize}
\item lengths of functions, which evaluates the capability of current approaches to handle different lengths of contexts;

\item different recommendation points, since different recommendation points affect how much context an approach can be aware of before recommendation;
\end{itemize}


\begin{figure}[t]
    \centering
    \includegraphics[width = 0.5\textwidth]{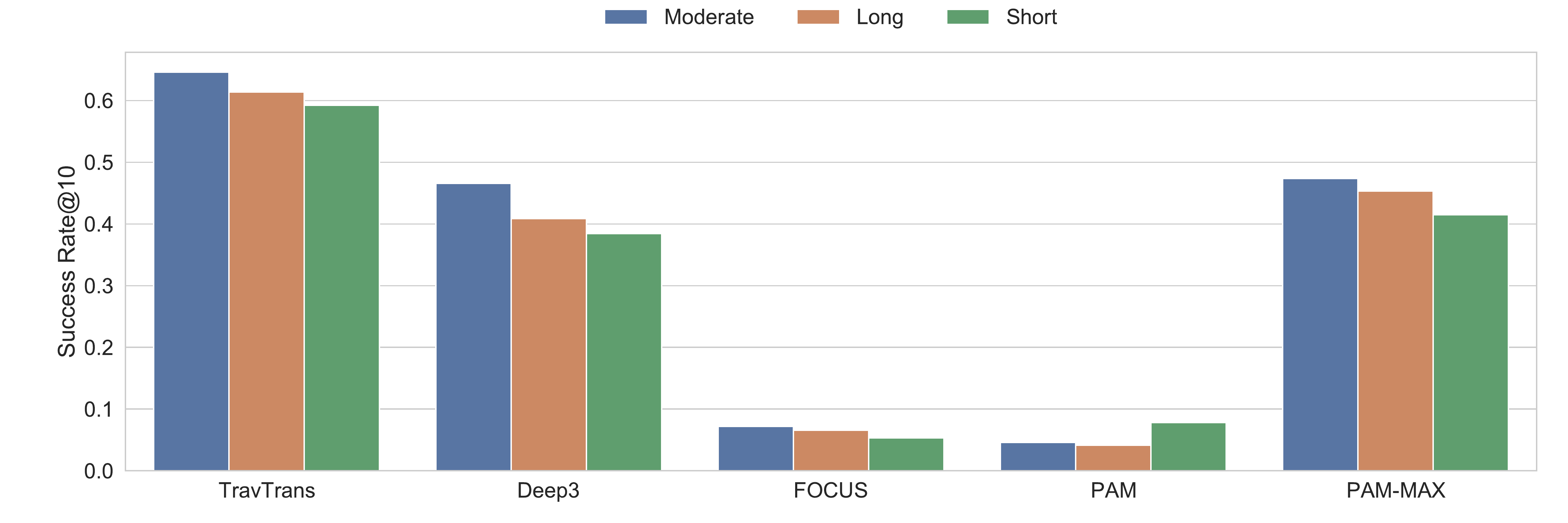}
    \caption{The Success Rate@10 of baselines on extremely short, normal and extremely long contexts at the ``{General}'' domain of \benchc.}
    \label{fig:length}
\end{figure}

\textbf{Capability to handle different lengths of functions.} In Section~\ref{sec:methodology} and Table~\ref{tab:benchc} we classified all functions of \benchc into extremely short functions, functions of moderate lengths, or extremely long functions by sampling the first 5\%, middle 90\% and last 5\% according to the distribution of function lengths. As code-based API recommendation is often based on the context in a function, the length of function can represent the length of context that an approach needs to handle. We study the performance of current baselines on functions of different lengths and show the results of TravTrans, Deep3, FOCUS, PAM, and PAM-MAX in Figure~\ref{fig:length}. 

From Figure~\ref{fig:length}, we find that most baselines share similar performance distributions on functions of different lengths. They present the best performance on functions with moderate lengths and suffer from performance drops on extremely long or short functions. To be more specific, the performance drops by 7.1\% for
extremely long functions and 10.6\% for extremely short functions on average. The results indicate that context length can affect the performance of current approaches. Besides, the approaches are more difficult to recommend correct APIs for the functions of extremely short lengths than those of extremely long lengths.

\finding{13}{Context length can impact the performance of current approaches in API recommendation. The approaches perform poorly for the functions with extremely short or long lengths, and accurate recommendation for the extremely short functions is more challenging.}

\textbf{Capability to handle different recommendation points.} Similar to the previous work ~\cite{nguyen2019focus}, we first define three locations of recommendation points. Suppose that the LOC of a function is $n$ and the total number of APIs used in the function is $m$. 
we define a recommendation point that is on the $a_{th}$ line of the function and is the $b_{th}$ API in the function locates on

1) the front of function if $a/n < 1/4$ and $b/m < 1/4$, or 

2) the middle of function if $1/4 < a/n < 3/4$ and $1/4 < b/m < 3/4$, or 

3) the back of function if $a/n > 3/4$ and $b/m > 3/4$.

\begin{figure}[t]
    \centering
    \includegraphics[width = 0.5\textwidth]{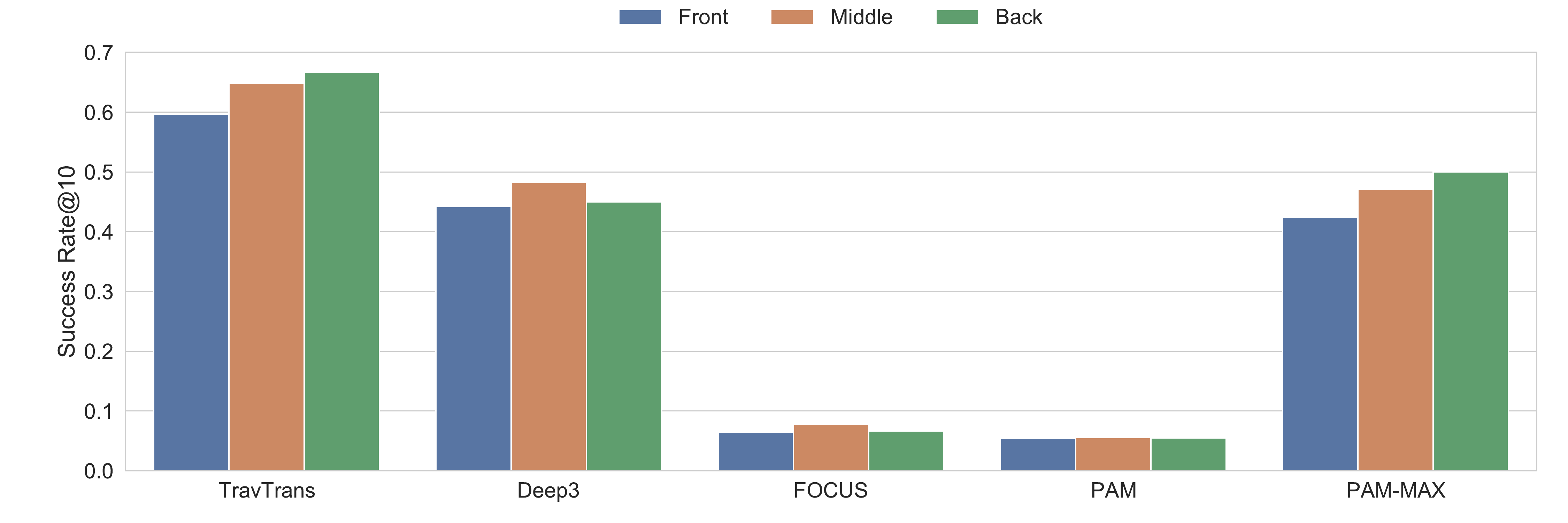}
    \caption{The Success Rate@10 of baselines on three categories of recommendation points at the \textit{general} domain of \benchc.}
    \label{fig:frontmiddleback}
\end{figure}

For all the APIs in the test set of the ``General'' domain, we replace them with placeholders of the above three types of recommendation points for evaluation. We also remove APIs in extremely long or short functions (according to the thresholds shown in Table~\ref{tab:benchc}) to alleviate the influence of function lengths. We show the results of TravTrans, Deep3, FOCUS, PAM and PAM-MAX in Figure~\ref{fig:frontmiddleback}. 

\begin{table*}[t]
    \centering
    \caption{The cross-domain Success Rate@10 of Python code-based API recommendation baselines. The rows list the domains where three baselines are trained and the columns list the domains where three baselines are evaluated. The \textcolor{red}{red} number indicates the best performance an approach achieves when trained on one domain (The largest number in each row). The numbers with \colorbox{mygray}{gray} background indicates the best performance achieved on a specific testing domain (The largest number in each column).}
    \begin{tabular}{cccccccccccccccc}
    \toprule
        \multirow{2}*{\textbf{Training Domain}} & & \multicolumn{4}{c}{\textbf{TravTrans}}  & & \multicolumn{4}{c}{\textbf{Deep3}} &  & \multicolumn{4}{c}{\textbf{PyART}}  \\
        \cmidrule{3-6}
        \cmidrule{8-11}
        \cmidrule{13-16}
        & & ML & Security & Web & DL & & ML & Security & Web & DL & & ML & Security & Web & DL \\
        \midrule
        ML & & 0.64 & 0.58 & 0.53 & \textcolor{red}{\textbf{0.71}} & & 0.42 & 0.41 & 0.36 & \textcolor{red}{\textbf{0.48}} & & 0.39 & 0.35 & 0.40 & \textcolor{red}{\textbf{0.40}}\\
        \specialrule{0em}{1pt}{1pt}
        Security & & 0.40 & \textcolor{red}{\textbf{0.54}} & 0.54 & 0.39 & & 0.31 & \textcolor{red}{\textbf{0.51}} & 0.42 & 0.29 & & 0.36 & \g \textcolor{red}{\textbf{0.48}} & 0.47 & 0.36\\
        \specialrule{0em}{1pt}{1pt}
        Web & & 0.54 & 0.63 & \textcolor{red}{\textbf{0.64}} & 0.51 & & 0.33 & 0.42 & \textcolor{red}{\textbf{0.46}} & 0.31 & & 0.42 & 0.47 & \g \textcolor{red}{\textbf{0.50}} & 0.40\\
        \specialrule{0em}{1pt}{1pt}
        DL& & 0.66 & 0.58 & 0.50 & \textcolor{red}{\textbf{0.68}} & & 0.44 & 0.39 & 0.33 & \textcolor{red}{\textbf{0.44}} & & 0.43 & 0.36 & 0.38 & \textcolor{red}{\textbf{0.45}}\\
        \midrule
        General & & \g 0.72 & \g 0.76 & \g 0.78 & \g 0.74 & & \g 0.55 & \g 0.65 & \g 0.62 & \g0.57 & & \g 0.44 & 0.44 & 0.46 & \g 0.46\\
    \bottomrule
    \end{tabular}
    \label{tab:pythondomainres}
\end{table*}

From Figure~\ref{fig:frontmiddleback}, we observe that most approaches obtain the best performance on the back recommendation points and acquire the worst performance on the front recommendation points. To be more specific, on average five approaches achieve 0.316, 0.347 and 0.348 for Success Rate@10 at the front, middle and back recommendation points, respectively. The results demonstrate that the current approaches tend to be less effective on the front recommendation points, which is reasonable since the context information is much limited before the front recommendation points.

\finding{14}{The location of recommendation points can affect the performance of current approaches. Current approaches are more effective in handling back recommendation points than front recommendation points.}

\eat{

\textbf{Impacts of class names.} He \etal~\cite{he2021pyart} finds that the type information can greatly help improve the performance of code-based API recommendation approaches in dynamic programming languages. Type information provides a explicit hint for recommendation approaches to narrow the search space into methods in a few classes since a type only supports a quite limited method set. Inspired by this finding, we generalize the setting into all programming languages by checking the presence of classes in current contexts. We give an example in listing.~\ref{lst:code}. In line 3 the API \textit{isfile} is called by declaring that it is from the class \textit{os.path} while in line 8 there is no declaration. When an approach needs to recommend this API, the contexts it sees are line 4 and line 9. Such cases exist both in static and dynamic languages. Therefore, it is more general and significant to study the impact of presence of class names on performance of current baselines.

\begin{figure}[t]
    \centering
    \includegraphics[width = 0.5\textwidth]{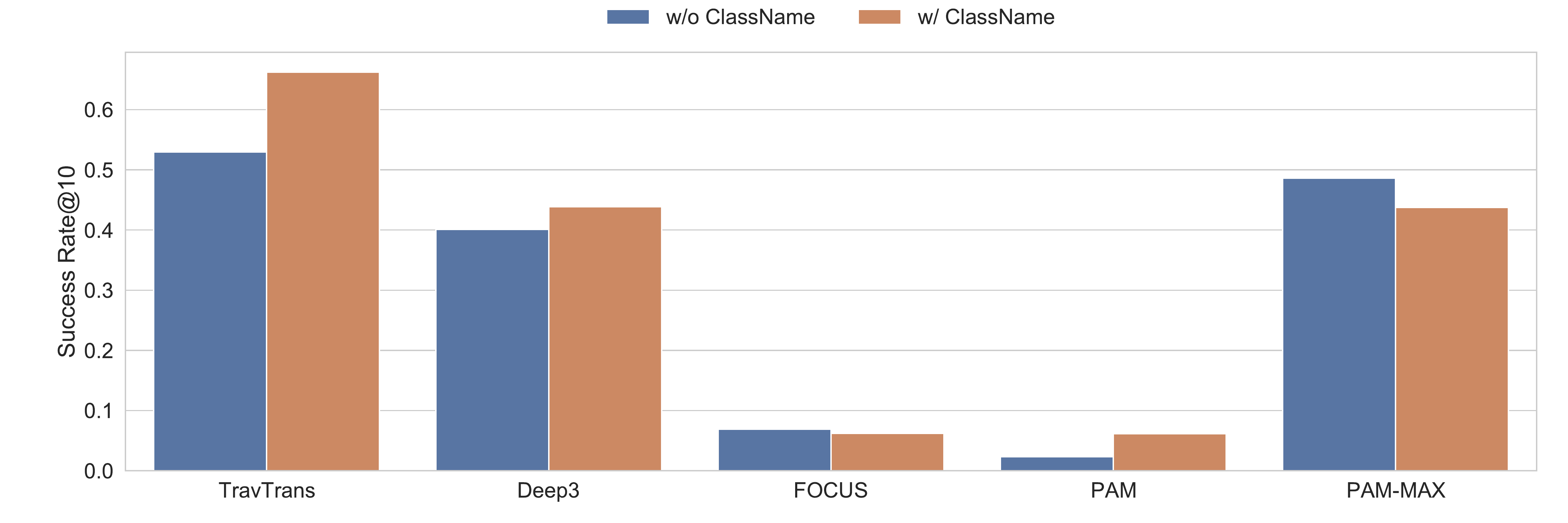}
    \caption{The Success Rate@10 of baselines based on contexts w/ and w/o class names at the \textit{general} domain of \benchc.}
    \label{fig:pureattr}
\end{figure}

\begin{lstlisting}[language = python, caption = The example of an API is called with and without its class name, label = lst:code]
#Called with corresponding class name
import os
os.path.isfile("a file")
os.path.<Recommendation Point>

#Called without corresponding class name
from os.path import isfile
isfile("a file")
<Recommendation Point>
\end{lstlisting}

To study the impact of class names, we divide the APIs in the testing set under \textit{general} domain of \benchc into two categories by checking whether they are called by classes or individually called. We test the performance of TravTrans, Deep3, FOCUS, PAM-MAX and PAM on two categories of APIs and show the results in figure.~\ref{fig:pureattr}.

\textcolor{red}{It is hard to describe the distribution without the data of FOCUS.}

\finding{15}{Class names in contexts play a important role in for approaches using token-based context since the performance of them drops by xxx $\sim$ xxx to handle contexts without class names.}

}

To sum up, different contexts can affect the performance of current code-based API recommendation approaches. Among them, the extremely short contexts and front recommendation points bring the most challenges for the accurate recommendation.

\subsection{Adaptation to Cross-Domain Projects (RQ6)}
We have divided \benchc into five different domains in Section~\ref{sec:methodology}. In this section, we aim at studying the adaption capability of current approaches for cross-domain projects. We train the approaches in one domain and evaluate them in other different domains. We choose the approaches TravTrans, Deep3, and PyART, which are all designed for Python, for analysis. We do not involve the approaches FOCUS, PAM, or PAM-MAX, since they use coarse-grained context representations or context-insensitive feature, and are difficult to incorporate project-specific information.
The first four rows of Table~\ref{tab:pythondomainres} list the cross-domain Success Rate@10 of TravTrans, Deep3 and PyART, respectively.

According to Table~\ref{tab:pythondomainres}, the approaches trained on one domain generally perform best on the test set of the same domain. For example, when trained on data from the ``Security'' domain, TravTrans, Deep3 and PyART obtain the best scores at 0.54, 0.51 and 0.48 on the test set of the same domain, respectively, in terms of Success Rate@10. However, their performance drops by 2.1\% $\sim$ 43.1\% when recommending APIs from different domains.

\finding{15}{Current approaches using fine-grained context representation are sensitive to the domain of the training data and suffer from performance drop when recommending cross-domain APIs.}

We also analyze the cross-domain performance of the approaches when training on multiple domains instead of on one single domain. Such analysis is worthwhile to explore whether different domains can complement each other. Then we train the approaches on the projects from the ``General'' domain of \benchc and evaluate them on the other four different domains.
We show the results in the last row of table~\ref{tab:pythondomainres}.

From the table, we can see that the approaches trained on the ``General'' domain generally show the best performance when evaluating on different domains. For example, TravTrans trained on the ``General'' domain achieves the Success Rate@10 of 0.72, 0.76, 0.78 and 0.74 on ML, Security, Web and DL domains, respectively, which is significantly higher than the corresponding best scores obtained by TravTrans trained on single one domain. We observe an average boost of 14\% for the performance of the approaches when trained on multiple domains than on a single domain. The results indicate that training approaches on multiple domains greatly improve the recommendation performance.

\finding{16}{Training on multiple domains helps the current approaches to recommend APIs in different
single domains, and the performance is generally
better than 
only training
on 
a single domain.}

\section{Discussion and Future Work}\label{sec:discussion}

\subsection{Query Reformulation for Query-based API Recommendation}

In Section~\ref{sec:rq2}, we find that query reformulation techniques can not only help current query-based API recommendation approaches find more correct APIs but also improve the ranking performance. Based on query reformulation, BIKER can even achieve a Success Rate@10 of 0.80 in class-level and 0.51 in method-level API recommendation. The results demonstrate that query quality has a great impact on the recommendation results and suggest that query reformulation should become a common pre-processing technique used before query-based API recommendation. 
We also discover that some query reformulation techniques, such as adding predicted API class names or relevant words, can improve the performance of query-based API recommendation approaches.
However, to the best of our knowledge, few studies have considered integrating these techniques, which could be one major reason that current approaches achieve limited performance.

By implementing a random deletion strategy, in Section~\ref{sec:del} we find that user queries usually contain noisy words, which can bias the recommendation results. We summarize three kinds of cases in which a query contains noisy words. However, there exists very little work that aims to detect and eliminate the irrelevant words for recommendation systems, which poses a great challenge for current approaches to be robust when handling various user queries. Although a random deletion strategy reduces the overall performance on average, the positive improvement of deletion on some specific words indicates the potential benefits of noisy word deletion.

\implication{1}{Current query-based API recommendation approaches should be integrated with query reformulation techniques to be more effective.}

\subsection{Data Sources for Query-based API Recommendation}

In Section~\ref{sec:rq1-1}, we point out that current query-based API recommendation approaches face the problem of building a comprehensive knowledge base due to the lack of enough data such as query-API pairs. In Section~\ref{sec:datasource}, we further discover that there is a semantic gap between user queries and descriptions from the official documentation. Both the lack of enough data for knowledge base creation and the semantic gap increase the difficulty of accurate API recommendation based on only official documentation. Such challenges can not be easily solved by improving learning-based models or pattern-based models. One effective way to mitigate the difficulty is to involve Stack Overflow posts, as analyzed in Section~\ref{sec:datasource}.
While Stack Overflow is only one type of data source, our analysis demonstrates that adding appropriate data sources can improve the performance of query-based API recommendation approaches.

\implication{2}{Apart from query reformulation, adding appropriate
data sources provides another solution to bridge the gap between queries and APIs.}

\subsection{Low Resource Setting in Query-based API Recommendation}
In Section~\ref{sec:rq1-1}, we find that current learning-based methods do not necessarily outperform traditional retrieval-based methods. We attribute the results to the limited data such as query-API pairs in the query-based API recommendation task, which is a low-resource scenario~\cite{hedderich2021survey,Diab2020lowresource}. We also discover that pre-trained models such as BERT show superior performance in query reformulation in Section~\ref{sec:rq2}. This indicates that current pre-trained models can implicitly mitigate the semantic gap between user queries and official descriptions of APIs. Future work is suggested to explore how to make the best use of pre-trained models for query-based API recommendation based on limited available data.

\implication{3}{Few-shot learning with powerful pre-trained models can be a solution to further improve the performance of query-based API recommendation.}

\subsection{User-defined APIs}

In Section~\ref{sec:rq4}, we find that current code-based API recommendation approaches, no matter pattern-based or learning-based models,
all face the challenge of recommending user-defined APIs. User-defined APIs have become the major bottleneck to further improve the performance of current code-based API recommendation approaches. However, as user-defined APIs usually do not appear in the training set, they can hardly
be learned by machine learning methods or be mined by pattern-based methods. A possible solution used by current approaches~\cite{huang2018api,facebooktransformer} is to regard the API as a code token and predict the token based on previous contexts. However, this solution also fails if the API token never appears in previous context. Thus, accurately predicting user-defined APIs should be one major direction of code-based API recommendation in future work.

\implication{4}{User-defined API recommendation is one major bottleneck for improving the performance of current code-based API recommendation approaches and remains unsolved.}




\eat{

\subsection{Uniform Scenario 1: \yun{Combination of} Code-based
\yun{and Query-based API recommendation for Code with Comments}
}
In this study, we clearly define the boundary between query-based and code-based API recommendation. Query-based API recommendation handles contexts written in natural language and code-based API recommendation handles contexts written in programming language. However, a natural question could be: can these two kinds of API recommendation exist in the same scenario? The answer is yes. Developers leverage comments to indicate the major function of the code and make the code more understandable. Comments are also written in natural language and could be regarded as a kind of queries. Therefore, a naive combination of query-based and code-based API recommendation could be using query-based approaches to handle the comments while using code-based approaches to handle code contexts and finally aligning the results of these two approaches. This combination makes sense but it ignores the possible relationship between comments and code. However, we believe this could be a good starting point to study the combination of query-based and code-based API recommendation to better help developers write code.

\implication{5}{Query-based and code-based API recommendation approaches can \yun{be combined together for handling the code with comments.}
}
}

\subsection{Query-based API Recommendation with Usage Patterns}

In this paper, we only focus on testing whether an approach can recommend the correct APIs, but we believe developers can always benefit more from detailed information about how to use the recommended APIs. A common method is to provide summaries such as the signature and constraints extracted from official documentation along with the recommended APIs. For example, KG-APISumm proposed by Liu \etal~\cite{liu2019generating} provides a detailed summary of the recommended API class. However, official documentation sometimes cannot provide enough usage information about an API, which may cause API misuse. For instance, a fresh developer may search ``\textit{how to read a file}'' in Python and the recommended API should be \textit{fileObject.read()}, but without sufficient experience to use file operations, the developers may forget to close the file after reading it. 

A possible solution to complement official documentation and avoid possible misuse is to provide usage patterns from other developers. In the above example, a common usage pattern \textit{open()}, \textit{fileObject.read()}, \textit{fileObject.close()} can prevent dangerous file operations. As there exist some pattern mining approaches on code, we can combine query-based API recommendation with code-based pattern mining methods for better providing the usage pattern.

\implication{5}{Code-based API recommendation approaches can provide usage patterns to enrich the results returned by query-based API recommendation approaches.}

\section{Threat To Validity}\label{sec:threat}

In this section, we describe the possible threats we may face in this study and discuss how we mitigate them.

\subsection{Internal Validity}

Our research may face the following internal threats:

\textbf{Baseline Re-implementation.} In this paper, we re-implemented several baselines according to the code or replication packages released by their authors. However, as some baselines are not primarily designed for API recommendation, we slightly modified their code and adapted them into our task and our benchmark. For example, we limit the prediction scope of code completion baselines to only API tokens. Such adaptations may cause the performance of baselines to be slightly different from the original papers. To mitigate this threat and validate the correctness of our re-implementation, we refer to some related work that cites these baselines and confirm our experiment results with them.

\textbf{Data Format.} In this paper, we align the performance of all code-based API recommendation baselines with a uniform evaluation process. However, different approaches may require different input formats and different pre-processing methods. For example, TravTrans requires the inputs of ASTs so we need to transform the code in \bench into ASTs. To mitigate the possible impacts brought by such transformations, we make sure that except for the original source code we do not involve extra information and knowledge in this modification process.

\textbf{Data Quality.} We build \benchq by manually selecting and labeling API-related queries from Stack Overflow and some tutorial websites. This process involved some human checks so that some subjective factors may influence the quality of our dataset. To mitigate this threat, we involve at least two persons to label one case and let one of our authors further check if the previous two persons give different opinions to the case. We also implement some rules to automatically filter out the cases that are explicitly unrelated to API recommendation.

\subsection{External Validity}

Our research may face the following external threats:

\textbf{Data Selection.} To the best of our knowledge, \bench is the largest benchmark in API recommendation task. We try to make it more representative by selecting real-world code repositories from the most popular domains at GitHub and real-world queries from the largest Q\&A forum StackOverflow according to several developer surveys~\cite{javasurvey, jetbrainssurvey}. All findings in this empirical study are based on this dataset. However, there may still be slight differences when adapting our findings into other domains and datasets that we do not discuss in this paper.

\textbf{Programming Language.} Our study focuses on the API recommendation on Python and Java, the findings included in this study may not be generalized to other programming languages that have different API call patterns from Python and Java. However, we believe the impacts of programming languages should not be significant as Python and Java are the most representative dynamically typed and statically typed languages, respectively, while most programming languages can be classified into these two categories.

\section{Related Work}\label{sec:relatedwork}
In this section, we list the work related to query reformulation, API recommendation and API-related empirical study, respectively.

\subsection{Query Reformulation}
The effectiveness of query-based approaches highly depends on the input natural language queries, which motivates many previous query reformulation works. Query expansion and query modification are two major reformulation methods, one of which adds several missing words and phrases which are relevant to the queries, the other one replaces or modifies words and phrases. Existing works show that even some minor textual changes influence the final results a lot. To better evaluate the API recommendation methods, we utilize multiple existing methods to reformulate the original queries. 

Rahman \etal propose ACER~\cite{mohammad2017improved}. It takes the initial query as the input and identifies appropriate search terms from source code by using a novel term weight named CodeRank, and then it suggests the best reformulation of the original query by using document structures, query quality analysis and machine learning techniques. 
Rahman \etal propose NLP2API~\cite{rahman2018nlp2api}, which automatically identifies the relevant API classes for a given query and then uses these API classes to reformulate the original query. 
Lu \etal propose to expand the queries by using synonyms. They extract the natural phrases from code identifiers, match the expanded queries with the identifiers, and then sort the methods in the code base by the matched identifier ~\cite{Lu2015queryexpansion}. 
Sirres \etal present COCABU~\cite{sirres2018augmenting} to resolve the vocabulary mismatch problem. Their approach uses common developer questions and expert answers to augment the user queries and improves the relevance of returned code examples. 
Nie \etal propose QECK~\cite{nie2016queryexpansion} to solve the term mismatch problem. QECK retrieves relevant question-answer pairs collected from Stack Overflow as the Pseudo Relevance Feedback (PRF) documents and identifies the specific words from the documents to expand the original query.
    
There are also some query reformulation techniques in industry or open-source communities. For example, the Google search engine utilizes its own prediction service~\cite{gps} to reformulate the natural language queries provided by users. The open-source library NLPAUG~\cite{ma2019nlpaug} provides NLP utilities to conduct data augmentation.

\subsection{API Recommendation}
\subsubsection{Query Based API Recommendation}
Multiple existing works ~\cite{mcmillan2011portfolio, zhang2011facg, zheng2011crosslib, chan2012searching, monperrus2012should, thung2013fr, rahman2016rack, gu2016deep,  xiong2018automating, huang2018api, xu2018mulapi, sun2019enabling, yuan2019api, qi2019finding, ling2019graph, liu2019generating, zhou2020BRAID, zhou2020BRAID} explore the possibility to provide developers with concrete API recommendation, using the natural language queries as input. Most of these works utilize open-source code bases, and some also use the knowledge in crowd-sourcing forums and wiki websites for augmentation. 

\textbf{Retrieval-based methods.} Portfolio~\cite{mcmillan2011portfolio}, proposed by McMillan \etal, recommends relevant APIs by utilizing several NLP techniques and indexing approaches with spreading activation network (SAN) algorithms as well as PageRank~\cite{brin1998pagerank}. 
Zhang \etal supplement the call graph with control flow analysis and design Flow-Augmented Call Graph (FACG) to utilize it for API recommendation~\cite{zhang2011facg}.
Chan \etal model API invocations as API graphs and design subgraph search algorithm to recommend APIs~\cite{chan2012searching}.
Rahman \etal collect the crowdsourced knowledge on Stack Overflow to extract keyword-API correlations and find several relevant API classes based on them~\cite{rahman2016rack}.
Huang \etal propose BIKER~\cite{huang2018api} to bridge the lexical gap and knowledge gap that previous approaches faced during API recommendation. BIKER obtains API candidates from Stack Overflow, and uses the similarity between queries and documentations as well as Stack Overflow posts to recommend API methods.
Liu \etal propose KG-APISumm~\cite{liu2019generating}, which is the first knowledge graph designed for API recommendation. KG-APISumm sorts the APIs through similarity calculation between queries and relevant parts of the constructed knowledge graph to recommend API classes.

Other than coding problems encountered by developers, there is another source of natural language functionality descriptions for APIs: feature requests from product managers or users. Thung \etal propose a method to recommend APIs based on the feature requests by learning from other modifications of the projects~\cite{thung2013fr}. Xu \etal propose MULAPI~\cite{xu2018mulapi}, which takes feature locations, project repositories and API libraries into consideration when recommending APIs.

\textbf{Learning-based methods.} DeepAPI~\cite{gu2016deep}, proposed by Gu \etal, is the first approach that combines deep learning with API recommendation. It reformulates the API recommendation task as a query-API translation problem and uses an RNN Encoder-Decoder model to recommend API sequences.
Xiong \etal propose to use representation learning to recommend web-based smart service~\cite{xiong2018automating}.
Ling \etal propose GeAPI~\cite{ling2019graph} based on graph embedding to provide more semantic information about and between APIs. GeAPI utilizes project's source code to automatically construct API graphs and leverages graph embedding techniques for API representation. Given a query, it searches relevant subgraphs on the original graph and recommends them to developers. 
Zhou \etal propose BRAID~\cite{zhou2020BRAID} and utilize approaches such as active learning as well as learning-to-rank based on the feedback of users to further improve the performance.

\subsubsection{Code Based API Recommendation}
\textbf{Pattern-based methods.}
Zhong \etal propose MAPO~\cite{zhong2009mapo} to mine API usage patterns and then recommends the relevant usage patterns to developers.
Sch{\"a}fer \etal propose Pythia~\cite{schafer2013effective} to utilize static pointer analysis and usage-based property inference to recommend APIs for JavaScript.
Wang \etal propose UP-Miner~\cite{wang2013mining} and use source code to extract succinct usage patterns to recommend APIs.
Nguyen \etal propose APIREC~\cite{nguyen2016api}, which uses fine-grained code changes and the corresponding changing contexts to recommend APIs.
D’Souza \etal propose PyReco~\cite{d2016collective}. It first extracts API usages from open-source projects and uses such information to rank the API recommendation results by utilizing nearest neighbor classifier techniques.
Fowkes \etal propose PAM~\cite{fowkes2016parameter} to tackle the problem that the recommended API lists are large and hard to understand. PAM mines API usage patterns through an almost parameter-free probabilistic algorithm and uses them to recommend APIs.
Niu \etal propose another API usage pattern mining approach, which segments the data using the co-existence relationship of object usages to mine API usage patterns~\cite{niu2017api}. 
Liu \etal propose RecRank~\cite{liu2018effective} to improve the top-1 accuracy based on API usage paths.
Nguyen \etal propose FOCUS~\cite{nguyen2019focus}, which mines open-source repositories and analyzes API usages in similar projects to recommend APIs and API usage patterns based on context-aware collaborative-filtering techniques.
Wen \etal propose FeaRS~\cite{wen2021siri}, which mines open-source repositories and extracts API sequences that are implemented together in the same tasks frequently to recommend APIs.

\textbf{Learning-based methods.}
Hindle \etal adopt the n-gram model, a widely-used statistical language model, on the code of software~\cite{hindle2016naturalness}, and develop a code suggestion tool based on the n-gram model.
Tu \etal propose to enhance the n-gram model by adding a cache component~\cite{tu2014localness}.
Raychev \etal propose to extract API sequences from open-source projects and index them into statistical language models to recommend APIs~\cite{raychev2014code}. 
Nguyen \etal propose  a graph-based language model GraLan~\cite{nguyen2015graph}. Based on GraLan, they design an AST-based language model named ASTLan to recommend APIs.
Raychev \etal propose a probabilistic model with decision trees named TGEN~\cite{Raychev2016prob} to predict code tokens.
Several recent works try to utilize syntax and data flow information for more accurate recommendation, besides focusing on token sequences~\cite{he2021pyart, kim2021code}.
He \etal propose PyART~\cite{he2021pyart}, which utilizes a predictive model along with data-flow, token similarity and token co-occurrence to recommend APIs.
Kim \etal leverage Transformer-based techniques to learn the syntactic information from source code~\cite{kim2021code}.

\subsection{Empirical Study on APIs}
There are several empirical studies focusing on different aspects of APIs.
Zhong \etal conduct an empirical study on how to use diversified kinds of APIs~\cite{zhong2019usages}.
Hora \etal study API evolution together with its influences on large software ecosystem~\cite{rw-api-evo-1}.
Zhong \etal study API parameter rules~\cite{rw-api-rule-1}.
Monperrus \etal conduct an empirical study on rules in API documentation~\cite{rw-api-rule-2}.
Ajam \etal study API topic issues on Stack Overflow~\cite{rw-api-so-1}, and Linares V{\'{a}}squez \etal focus on how API changes affect activities on Stack Overflow~\cite{rw-api-so-2}.
Moreover, other work with respect to cross-language API mapping relations~\cite{rw-api-mapping}, API migration~\cite{rw-api-migration}, API usability~\cite{rw-api-usability}, API deprecation~\cite{rw-api-deprecation}, API learning obstacles~\cite{rw-api-obstacle}, usage of specific types of APIs~\cite{rw-api-parallel, rw-api-trivial}, the correlation between APIs and software quality~\cite{rw-api-proneness}, as well as order of API usage patterns~\cite{rw-api-order}, are also explored.

\section{Conclusion}
In this paper, we present an empirical study on the API recommendation task. We classify current work into query-based and code-based API recommendation, and build a benchmark named \bench to align the performance of different recommendation approaches. We conclude some findings based on the empirical results of current approaches.

For query-based API recommendation approaches, we find that 1) recommending method-level APIs is still challenging; 2) query reformulation techniques have great potential to improve the quality of user queries thus they can help current approaches better recommend APIs. What's more, user queries also contain some meaningless and verbose words and even a simple word deletion method can improve the performance; 3) approaches built upon different data sources have quite different performances. Q\&A forums such as Stack Overflow can greatly help mitigate the gap between user queries and API descriptions. 

For code-based API recommendation, we emphasize the superior performance of current deep learning models such as Transformer. However, they still face the challenge of recommending user-defined APIs. We also find different contexts, such as different location of recommendation points and context length, can impact the performance of current approaches. Besides, current approaches suffer from recommending cross-domain APIs.

Based on the findings, we summarize some future directions on improving the performance of API recommendation. For query-based approaches, we encourage to integrate query reformulation techniques with query-based API recommendation approaches to obtain better performance, but how to choose the best query reformulation strategy still remains as future work. We also believe some few-shot learning methods and different data sources can bridge the gap between user queries and knowledge base under low resource scenarios. For code-based approaches, we recommend future work to focus on improving the performance of user-defined API recommendation and train the approach on multiple domains instead of a single domain.

We released our benchmark \bench and all experiment results at Github\footnote{The address of benchmark is at \url{https://github.com/JohnnyPeng18/APIBench}}. We hope this empirical study can remove some barriers and motivate future research on API recommendation.






%

\newpage

\bibliographystyle{plain}
\bibliography{ref}

\end{document}